%% file: main.tex
\documentclass[10pt, conference, compsocconf]{IEEEtran}
\ifCLASSINFOpdf
\else
\fi

\let\labelindent\relax
\usepackage{cite}
\usepackage{amsmath,amssymb,amsfonts}
\usepackage{graphicx}
\usepackage{textcomp}
\usepackage{bbding}
\usepackage{pifont}
\usepackage{multicol}

\usepackage{booktabs}
\usepackage[binary-units=true]{siunitx}
\usepackage{array,multirow}
\usepackage{hyperref}
\usepackage{amsmath}
\usepackage{enumitem}
\usepackage{algorithm}
\usepackage{graphicx}
\usepackage{float}
\usepackage[symbol, hang,flushmargin]{footmisc}
\usepackage{bm}

\usepackage{tablefootnote}

\usepackage[noend]{algpseudocode}

\newcommand{\figref}[1]{Figure~\ref{#1}}
\newcommand{\secref}[1]{Section~\ref{#1}}

\newcommand{\tabref}[1]{Table~\ref{#1}}

\newcolumntype{L}[1]{>{\raggedright\let\newline\\\arraybackslash\hspace{0pt}}m{#1}}
\newcolumntype{C}[1]{>{\centering\let\newline\\\arraybackslash\hspace{0pt}}m{#1}}
\newcolumntype{R}[1]{>{\raggedleft\let\newline\\\arraybackslash\hspace{0pt}}m{#1}}

\usepackage{soul}
\usepackage[dvipsnames]{xcolor}
\usepackage{xurl}

\definecolor{light_red}{RGB}{255, 204, 204}
\definecolor{crimson}{rgb}{0.86, 0.08, 0.24}    

\newif\ifcomment

\commentfalse

\ifcomment
\newcommand{\hongxiang}[1]{\sethlcolor{yellow}\hl{[Hongxiang: #1]}}
\newcommand{\thomas}[1]{\sethlcolor{magenta}\hl{[Thomas: #1]}}
\newcommand{\stelios}[1]{\sethlcolor{green}\hl{[Stelios: #1]}}
\newcommand{\alex}[1]{\sethlcolor{cyan}\hl{[Alex: #1]}}
\newcommand{\royson}[1]{\sethlcolor{purple}\hl{[Royson: #1]}}
\newcommand{\mohamed}[1]{\sethlcolor{orange}\hl{[Mohamed: #1]}}
\newcommand{\cut}[1]{\sethlcolor{light_red}\hl{[#1]}}

\else
\newcommand{\hongxiang}[1]{}
\newcommand{\thomas}[1]{}
\newcommand{\stelios}[1]{}
\newcommand{\alex}[1]{}
\newcommand{\royson}[1]{}
\newcommand{\mohamed}[1]{}
\newcommand{\cut}[1]{}

\fi

\newif\ifarxiv
\arxivtrue
\usepackage{tikz}

\usepackage{lastpage}
\usepackage{fancyhdr}
\fancypagestyle{firstpage}
{
    \fancyhead{}
    \begin{tikzpicture}[remember picture,overlay]
    \node [xshift=153mm,yshift=-10mm]
    at (current page.north west) {\href{https://www.acm.org/publications/policies/artifact-review-and-badging-current}{\includegraphics[width=1.8cm]{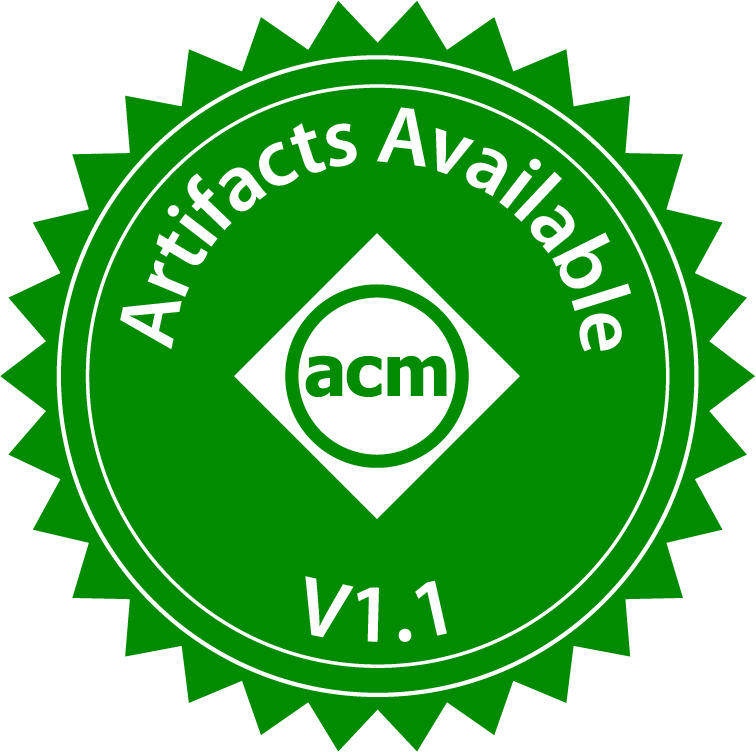}}} ;
    \node [xshift=172mm,yshift=-10mm]
    at (current page.north west) {\href{https://www.acm.org/publications/policies/artifact-review-and-badging-current}{\includegraphics[width=1.8cm]{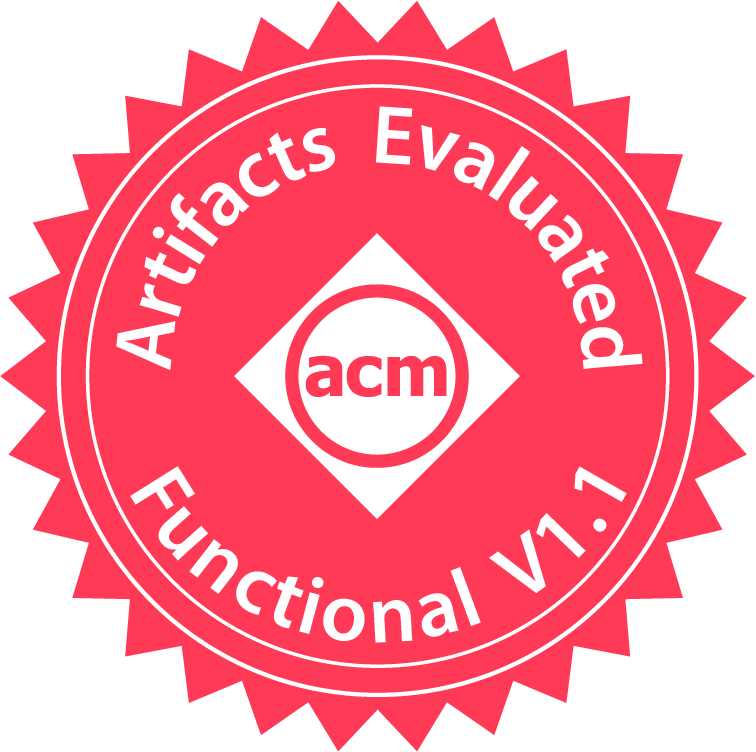}}} ;
    \node [xshift=191mm,yshift=-10mm]
    at (current page.north west) {\href{https://www.acm.org/publications/policies/artifact-review-and-badging-current}{\includegraphics[width=1.8cm]{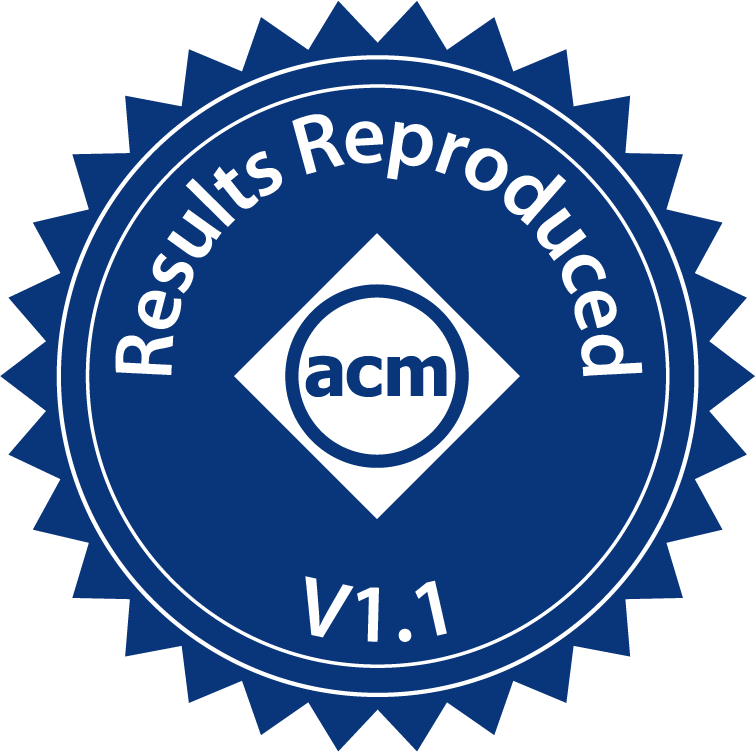}}} ;
    \end{tikzpicture}

  \ifarxiv
  \fancyfoot[C]{\scriptsize \copyright~2022 IEEE. Personal use of this material is permitted. Permission from IEEE must be obtained for all other uses, in any current or future media, including reprinting/republishing this material for advertising or promotional purposes, creating new collective works, for resale or redistribution to servers or lists, or reuse of any copyrighted component of this work in other works. 
  This work has been accepted at the IEEE/ACM International Symposium on Microarchitecture (MICRO’22).}
  \else
    
  \fi
}

\makeatletter
\def\ps@IEEEtitlepagestyle{%
  \def\@oddfoot{\mycopyrightnotice}%
  \def\@oddhead{\hbox{}\@IEEEheaderstyle\leftmark\hfil\thepage}\relax
  \def\@evenhead{\@IEEEheaderstyle\thepage\hfil\leftmark\hbox{}}\relax
  \def\@evenfoot{}%
}
\def\mycopyrightnotice{%
  \begin{minipage}{\textwidth}
  \scriptsize
  \copyright~2021 IEEE. Personal use of this material is permitted. Permission from IEEE must be obtained for all other uses, in any current or future media, including reprinting/republishing this material for advertising or promotional purposes, creating new collective works, for resale or redistribution to servers or lists, or reuse of any copyrighted component of this work in other works. 
  
  This work has been accepted at The International Conference on Field-Programmable Technology (FPT’21).
  \end{minipage}
}
\makeatother

\makeatletter
\newcommand{\linebreakand}{%
  \end{@IEEEauthorhalign}
  \hfill\mbox{}\par
  \mbox{}\hfill\begin{@IEEEauthorhalign}
}
\makeatother

\begin{document}
\title{Adaptable Butterfly Accelerator for Attention-based NNs via \\ Hardware and Algorithm Co-design}


\author{\IEEEauthorblockN{Hongxiang Fan}
\IEEEauthorblockA{ Imperial College London \\
London, UK\\
h.fan17@imperial.ac.uk
}
\and
\IEEEauthorblockN{Thomas Chau}
\IEEEauthorblockA{
Samsung AI Center \\
Cambridge, UK\\
thomas.chau@samsung.com}
\and
\IEEEauthorblockN{Stylianos I. Venieris}
\IEEEauthorblockA{
Samsung AI Center \\
Cambridge, UK\\
s.venieris@samsung.com}
\and
\IEEEauthorblockN{Royson Lee}
\IEEEauthorblockA{
University of Cambridge \\
Cambridge, UK\\
dsrl2@cam.ac.uk}
\linebreakand
\IEEEauthorblockN{Alexandros Kouris}
\IEEEauthorblockA{
Samsung AI Center \\
Cambridge, UK\\
a.kouris@samsung.com}
\and
\IEEEauthorblockN{Wayne Luk}
\IEEEauthorblockA{ Imperial College London \\
London, UK\\
w.luk@imperial.ac.uk
}
\and
\IEEEauthorblockN{Nicholas D. Lane}
\IEEEauthorblockA{ 
Samsung AI Center \&  \\
University of Cambridge \\
Cambridge, UK\\
nic.lane@samsung.com
}
\and
\IEEEauthorblockN{Mohamed S. Abdelfattah}
\IEEEauthorblockA{ Cornell University \\
New York, NY, United States\\
mohamed@cornell.edu
}
}


%Telephone: (800) 555--1212, Fax: (888) 555--1212}


\maketitle
\thispagestyle{firstpage}

\begin{abstract}
Attention-based neural networks have become pervasive in many AI tasks.
Despite their excellent algorithmic performance,
the use of the attention mechanism and feed-forward network (FFN) demands excessive computational and memory resources,
which often compromises their hardware performance.
Although various sparse variants have been introduced,
most approaches only focus on mitigating the quadratic scaling of attention on the algorithm level, without  explicitly considering the efficiency of mapping their methods on real hardware designs.
Furthermore, most efforts only focus on either the attention mechanism or the FFNs but without jointly optimizing both parts, causing most of the
current designs to lack scalability when dealing with different input lengths.
This paper systematically considers the sparsity patterns in different variants from a hardware perspective.
On the algorithmic level, we propose \textit{FABNet}, a hardware-friendly variant that adopts a unified butterfly sparsity pattern to approximate both the attention mechanism and the FFNs. 
On the hardware level, a novel adaptable butterfly accelerator is proposed that can be configured at runtime via dedicated hardware control to accelerate different butterfly layers using a single unified hardware engine.
On the Long-Range-Arena dataset,
\textit{FABNet} achieves the same accuracy as the vanilla \textit{Transformer} while reducing the amount of computation by $10\sim66\times$ and the number of parameters $2\sim22\times$.
By jointly optimizing the algorithm and hardware,
our FPGA-based butterfly accelerator achieves {$14.2\sim23.2\times$} speedup over state-of-the-art accelerators normalized to the same computational budget.
Compared with optimized CPU and GPU designs on Raspberry Pi 4 and Jetson Nano,
our system is up to $273.8\times$ and $15.1\times$ faster under the same power budget.

\end{abstract}


% \fi
\IEEEpeerreviewmaketitle

\input{1_introduction}
\input{2_background}

\input{3_algorithm_opt}

\input{4_hardware_design}
\input{5_hardware_opt}

\input{6_evaluation}
\input{7_related_work}
\input{8_conclusion}
\input{9_artifact_evaluation}
\bibliographystyle{IEEEtran}
\bibliography{refs}

\end{document}

%% file: 1_introduction.tex
\section{Introduction}

\begin{figure}[t]\centering
\includegraphics[width=0.48\textwidth]{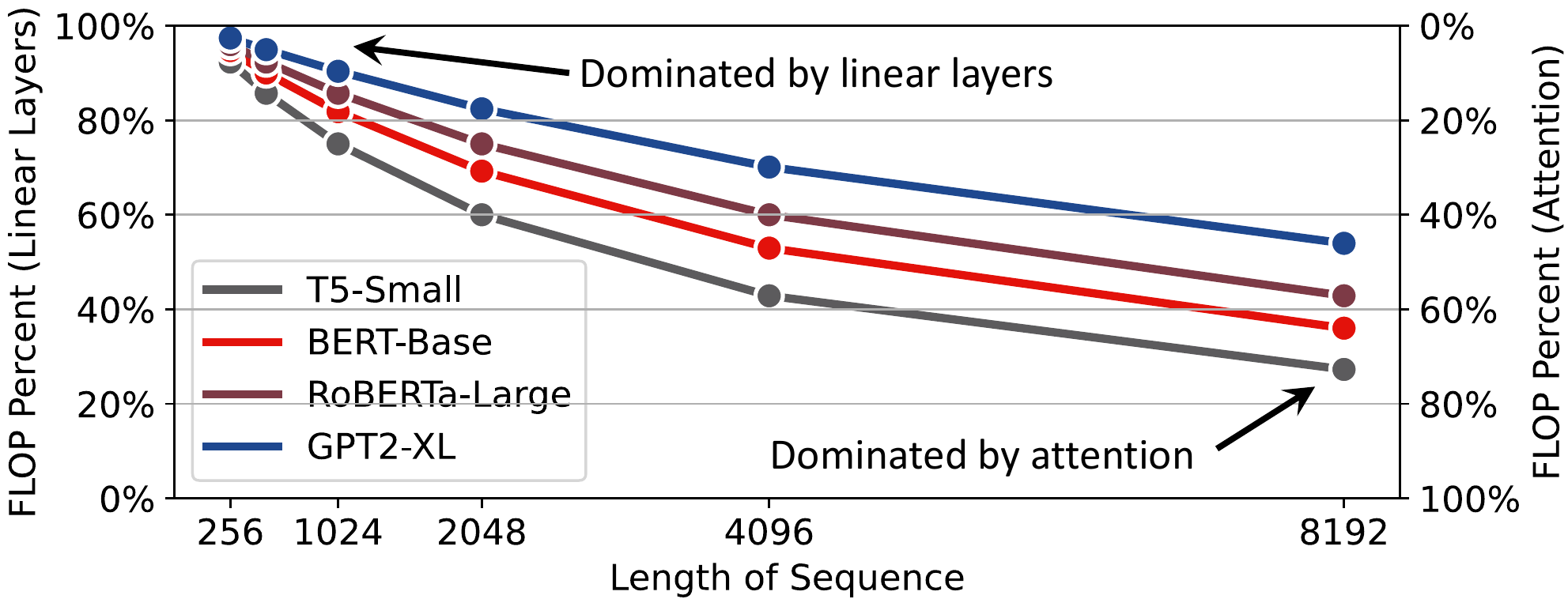}
\vspace{-2.0mm}
\caption{FLOPs percentage of attention and linear layers with different input sequence length.}\label{fig:op_count}
\vspace{-2.0mm}
\end{figure}

Recent years have witnessed a great success of attention-based neural networks (NNs) on many AI tasks~\cite{brauwers2021general}.
The attention mechanism~\cite{vaswani2017attention} that captures long-range information from sequences of data has demonstrated its excellent algorithmic performance in various natural language processing~\cite{devlin2018bert, radford2019language} and computer vision~\cite{dosovitskiy2020image} applications.
However, the advances of attention-based NNs come at a cost: the use of attention and linear layers significantly increases the computational load, resulting in a large overhead on their speed and power consumption~\cite{wang2020spatten}.
\figref{fig:op_count} shows an operation breakdown on four mainstream attention-based models.
For short input sequences, linear layers occupy over $80$\% of operation counts.
As the input sequence increases, the computation is gradually dominated by the attention layer.
Since both attention and linear layers are memory- and compute-intensive,
it is challenging to achieve high hardware performance on attention-based NNs across input sequences of various lengths.

So far, various approaches and designs have been introduced to accelerate attention-based DNNs.
On the algorithmic level, several efficient sparse variants have attempted to reduce the computational complexity~\cite{choromanski2020rethinking, wang2020linformer, kitaev2020reformer, tay2020sparse, beltagy2020longformer, zaheer2020big, child2019generating}.
However, most of these approaches only focus on reducing the number of parameters and operations without considering the real hardware performance, such as end-to-end latency.
Furthermore, the hardware efficiency of implementing these sparsity patterns on real hardware designs is often overlooked.
On the hardware level,
although various highly-optimized accelerators (\tabref{tb:sota_sparse}) have been proposed~\cite{ham2020, wang2020spatten, sanger201micro, zhou2021energon, elsa2021isca, dota2022asplos, li2020ftrans, edgebert2021micro}, several issues still remain unresolved:

\mohamed{the following list should be converted to a paragraph and the criticism toned down a bit}

\begin{itemize}[leftmargin=*]
    \item Most of current accelerators only focus on optimizing either FFNs \cite{li2020ftrans} or the attention mechanism~\cite{ham2020, wang2020spatten, sanger201micro, zhou2021energon, elsa2021isca, dota2022asplos}. Without jointly optimizing both parts, these hardware designs lack scalability when accelerating the end-to-end attention-based NNs with different input lengths.
    \item While optimizing the attention mechanism, most of the existing designs dynamically detect and prune the redundant computation at runtime to achieve high sparsity on specific datasets and networks.
    However, the generality of these dynamic approaches needs to be further tested as their performance gain may vary across different datasets and network architectures.%of these designs cannot be guaranteed since the sparsity
    \item The sparsity patterns introduced by these dynamic approaches are often unstructured, requiring dynamic hardware controllers to exploit sparsity.
    Such complicated controllers often contain larger numbers of clocking elements, and their hardware overhead increases as the transistor size reduces~\cite{jang2021sparsity}. As such, the performance or energy gain of these dynamic methods may be diminished.
\end{itemize}

\mohamed{We should be prepared to defend why specifically our optimizations are the right ones to do in hardware? Why not low-rank factorization for example? Can we maybe have a ROUGH estimate of the efficiency of low-rank in hardware? Is it the case that we have a smaller number of FLOPs in the general case, and we would be able to translate low flops to speedup if we build dedicated hardware? I think we need a motivating story like that.}

To address the aforementioned issues,
this paper adopts butterfly sparsity to accelerate attention-based models with three novel aspects~(\tabref{tb:sota_sparse}):
\textit{i)} fine-grained structured regularity, which possesses regular data accesses to optimize both memory and compute efficiency;
\textit{ii)} static sparsity pattern, which avoids the need of designing a dynamic controller in hardware;
\textit{iii)} sparsity exploitation on both attention and linear layers,
which allows scalable end-to-end acceleration of attention-based NNs.
We therefore propose \textit{FABNet}, a hardware-friendly model for \textbf{F}FT, \textbf{A}ttention and \textbf{B}utterfly-Net. To fully exploit the sparsity in hardware, 
we propose an adaptable butterfly accelerator that can be configured at runtime via dedicated hardware control to accelerate different layers using a single unified engine, significantly improving hardware efficiency.
To push the performance limit, we jointly optimize the model and hardware via a co-design approach.
Overall, this work makes the following contributions:

\begin{itemize}[leftmargin=*]
  \item A hardware-friendly attention-based model, \textit{FABNet}, that adopts the butterfly sparsity pattern on both attention and linear layers for end-to-end acceleration (\secref{sec:alg_opt}).
  \item A novel adaptable butterfly accelerator configurable at runtime via dedicated hardware control to accelerate different layers using a single unified engine (\secref{sec:hw}).
  \item Several hardware optimizations to improve the hardware efficiency and a co-design approach to jointly optimize both algorithmic and hardware parameters (\secref{sec:hw_opt}).
  \item A comprehensive evaluation on different datasets that demonstrates the advantages of our approach over CPU, GPU and state-of-the-art accelerators (\secref{sec:evaluation}).
\end{itemize}

\begin{table}[t]
\centering
\caption{Comparison of existing accelerators for attention-based NNs in terms of sparsity regularity, pattern and location. }
\label{tb:sota_sparse}
\setlength\tabcolsep{1pt}
\scalebox{0.8}{
\begin{tabular}{C{2.8cm}|C{3.99cm}|C{1.8cm}|C{1.8cm}}
\toprule
\textbf{Accelerators} & \textbf{Pattern Regularity} & \textbf{Sparsity Pattern} & \textbf{Sparsity Location} \\ \midrule 
$A^{3}$~\cite{ham2020} & unstructured & \multirow{8}{*}{dynamic} & \multirow{8}{*}{attention} \\ \cmidrule{1-2}
SpAtten~\cite{wang2020spatten} & coarse-grained structured & & \\ \cmidrule{1-2}
Sanger~\cite{sanger201micro} & load-balanced unstructured & & \\ \cmidrule{1-2}
Energon~\cite{zhou2021energon} & unstructured & & \\ \cmidrule{1-2}
ELSA~\cite{elsa2021isca} & unstructured & & \\ \cmidrule{1-2}
DOTA~\cite{dota2022asplos}&  unstructured & & \\ \midrule
FTRANS~\cite{li2020ftrans} & None & static & FFN \\ \midrule
EdgeBERT~\cite{edgebert2021micro} & None & dynamic & layer \\ 
\midrule
Our work & fine-grained structured & static & attention \& FFN \\ 
\bottomrule
\end{tabular}
}
\end{table}

%% file: 2_background.tex
\section{Background and Motivation}\label{sec:backgroud}

\subsection{Attention-Based Neural Networks}
\label{sec:attention_nns}
\begin{figure}[t]\centering
\includegraphics[width=0.49\textwidth]{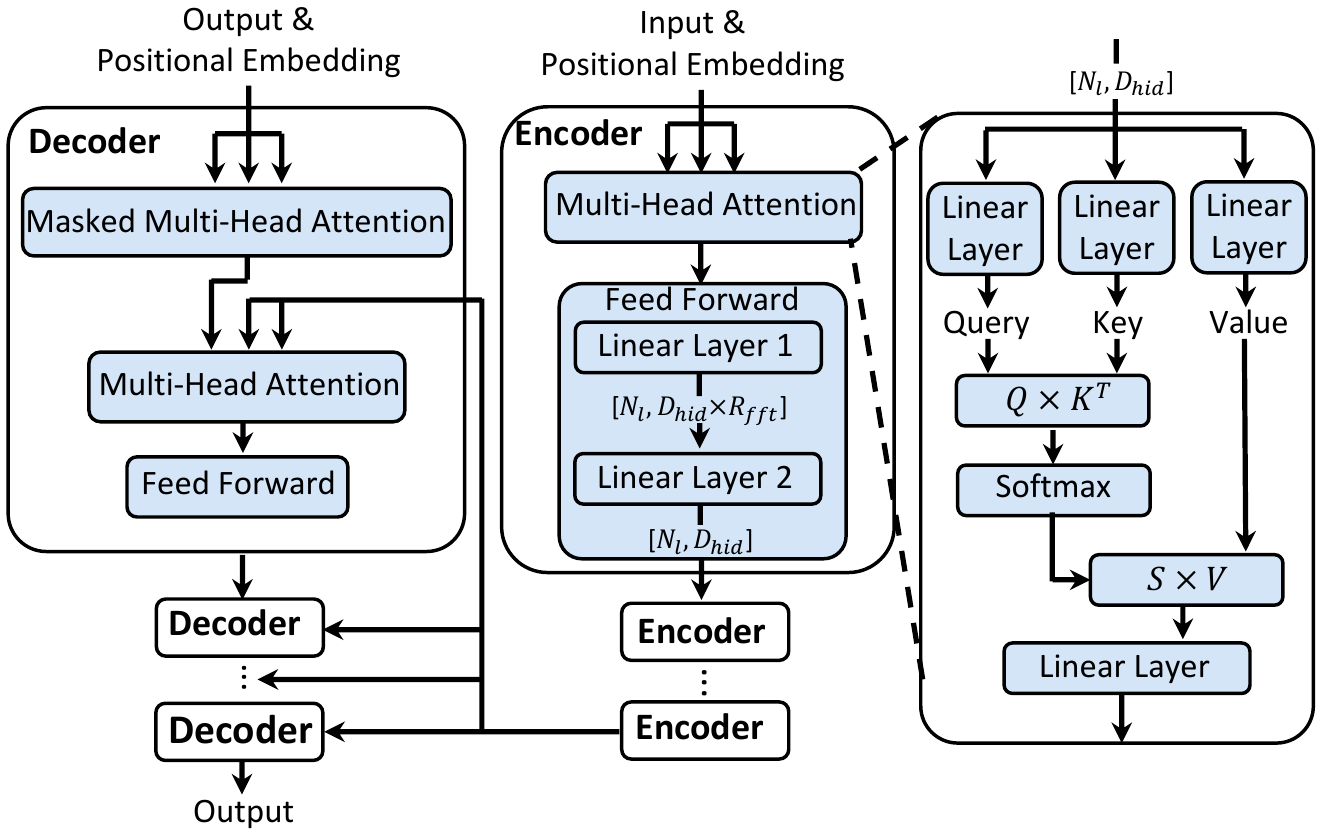}
\vspace{-5.0mm}
% \vspace{-6.0mm}
\caption{The structure of a {\textit{Transformer}}. Shortcut addition and layer normalization are omitted for simplicity.}\label{fig:transformer}
\vspace{-3.0mm}
\end{figure}

Based on their network structure, attention-based NNs can be classified into three categories: \textit{i)}~encoder-decoder, \textit{ii)}~encoder-only, and \textit{iii)}~decoder-only networks.
The encoder-decoder NNs are mainly designed for sequence-to-sequence tasks, such as machine translation~\cite{vaswani2017attention}.
One of the most widely used encoder-decoder network is the {\textit{Transformer}}, which is constructed by a stack of encoder and decoder blocks.
\figref{fig:transformer} illustrates the structure,
where $N_\text{l}$, $D_\text{hid}$ and $R_\text{fft}$ represent input length, hidden size and FFN expand ratio respectively.
Each encoder starts with a multi-head attention module, followed by a feed-forward network (FFN) consisting of two linear (fully-connected) layers.
Finally, residual addition~\cite{he2016deep} and layer normalization ({LN})~\cite{ba2016layer} are used after FFN.
Within each multi-head attention,
the inputs are first mapped to query ($\mathbf{Q}$), key ($\mathbf{K}$) and value ($\mathbf{V}$) matrices through three different linear layers.
The query matrix is then multiplied with $\mathbf{K}^{T}$,
followed by a softmax operation to get the score matrix ($\mathbf{S}$).
The generated $\mathbf{S}$ is multiplied with $\mathbf{V}$
and the resultant matrix will flow into another linear layer,
which generates the final output matrix of the multi-head attention.
Similar to the encoder,
the decoder employs two multi-head attention modules and one FFN, where the difference is that the inputs of the query and key matrices in the second attention module come from the last encoder. 

Based on the original encoder-decoder structure of {\textit{Transformer}}, different variants have been proposed.
The encoder-only networks, such as {BERT}~\cite{devlin2018bert} and {XLM}~\cite{lample2019cross}, are autoencoding models that have been widely applied to NLP tasks, such as sequence classification~\cite{wang2018glue}.
The Vision Transformer (ViT)~\cite{dosovitskiy2020image} also lies in this category. An extra linear projection layer is introduced at the beginning, while its encoder layers correspond to the encoder part of the original {\textit{Transformer}}.
Finally, the decoder-only networks represent the autoregressive models designed for NLP tasks, such as language modeling~\cite{ma2019tensorized}.
{GPT}~\cite{radford2019language} is a typical decoder-only model {that corresponds to} the decoder part of the original {\textit{Transformer}}.
Although we focus on encoder-only networks in this work, our hardware design is flexible and applicable to decoders too.

\subsection{Butterfly Matrices and FFT}\label{subsec:backgroud_bfly}
Despite the impressive accuracy attained using attention-based NNs, these models are expensive and not scalable, \textit{e.g.}~the self-attention mechanism in the \textit{Transformer} scales quadratically in terms of computation and memory as a function of the input sequence length.
As a result, numerous works~\cite{choromanski2020rethinking, wang2020linformer,child2019generating,chen2021scatterbrain} adopt structured linear mappings, such as sparse and low-rank matrices, to approximate the attention matrices and/or the weight matrices in the feed-forward layers.
Choosing an appropriate structure for each linear mapping, however, is application-dependent, often requiring domain expertise and entailing an arduous process of hand-picking solutions as different structures have different trade-offs in accuracy and speed.

To counteract this, recent work has utilized butterfly matrices~\cite{parker1995random,dao2019learning}, which are universal representations of structured matrices that have a simple recursive structure.
Specifically, each butterfly matrix \scalebox{0.8}{$\mathbf{W}_{\text{Bfly}}$} of size $N$ encodes the recursive divide-and-conquer structure with butterfly patterns
% of the Fast Fourier Transform (FFT) 
and, hence it can be expressed as the product of sparse butterfly factor matrices~\cite{de2018two} as follows:
% The butterfly matrix \scalebox{0.8}{$\mathbf{W}_{\text{Bfly}}$} can be mathematically formulated as:
\small
\begin{equation*}
\scalebox{0.8}{$\mathbf{W}_{\text{Bfly}}$} = 
    \left(\mathbf{W}^{'}_{N}
    \left[\begin{array}{cc}
    \mathbf{W}^{'}_{N / 2} & 0 \\
    0 & \mathbf{W}^{'}_{N / 2}
    \end{array}\right]
    \ldots\left[\begin{array}{ccc}
    \mathbf{W}^{'}_{2} & \ldots & 0 \\
    \vdots & \ddots & \vdots \\
    0 & \ldots & \mathbf{W}^{'}_{2}
    \end{array}\right]\right)
\end{equation*}
\normalsize
where each $\mathbf{W}^{'}_{N}$, a butterfly factor, is a $2\times2$ block matrix %of size $N$,
of diagonal matrices, $\mathbf{B}^{i}$ with size $N/2$, whose entries can be trained via gradient-based methods:
\small
\begin{equation*}
    \mathbf{W}^{'}_{N} = 
    \left[\begin{array}{cc}
    \mathbf{B}^{1}_{N / 2} & \mathbf{B}^{2}_{N / 2} \\
    \mathbf{B}^{3}_{N / 2} & \mathbf{B}^{4}_{N / 2}
    \end{array}\right].
\end{equation*}
\normalsize
Due to their expressiveness in representing structured matrices and approximating unstructured data, butterfly matrices and their variants~\cite{chen2021pixelated,dao2020kaleidoscope} have found success in compressing attention and weight matrices, considerably improving the accuracy and efficiency of attention-based NNs.
For instance, applying butterfly factorization to a linear layer with an $M\times M$ weight matrix can reduce the computational and memory complexity from $\mathcal{O}(M^2)$ to $\mathcal{O}(M\log{}M)$.

Besides attention and weight matrices, some designs have explored replacing the entire attention mechanism with more efficient counterparts~\cite{tolstikhin2021mlp}. A prominent example is FNet~\cite{lee2021fnet}, in which the self-attention modules are replaced by 2D Discrete Fourier Transform (DFT) operations.
Specifically, for each input,  1D DFT is applied along the sequence and the hidden dimension independently, keeping only the real component of the resulting outputs.
To reduce DFT computation time, the Cooley-Tukey Fast Fourier Transform (FFT) algorithm~\cite{cooley1965algorithm} is used.
As the use of DFT facilitates information flow across all embeddings, it results in a similar performance compared to the use of vanilla self-attention layers, but at a significant reduction in latency and memory.

On the algorithmic front, our proposed \textit{FABNet} utilizes a mixture of these techniques -- FFT and butterfly matrices -- to outperform relevant approximation approaches in terms of accuracy. Notably, since FFT matrices can be considered a special case of butterfly matrices with \scalebox{0.8}{$\mathbf{B}^{1}_{N / 2} $}, \scalebox{0.8}{$\mathbf{B}^{3}_{N / 2} $} being identity matrices and \scalebox{0.8}{$\mathbf{B}^{1}_{N / 2}$}, \scalebox{0.8}{$\mathbf{B}^{4}_{N / 2}$} acting as twiddle factors, both the FFT and butterfly matrices possess the recursive butterfly structure.
Therefore, it is possible to use a unified computational and data access pattern and then devise a single hardware engine to accelerate both FFT and butterfly-based operations with high hardware efficiency.

\subsection{Latency Breakdown and Motivation}\label{subsec:motivation}

\begin{figure}[t]\centering
\includegraphics[width=0.49\textwidth,trim={0 1cm 0 1cm},clip]{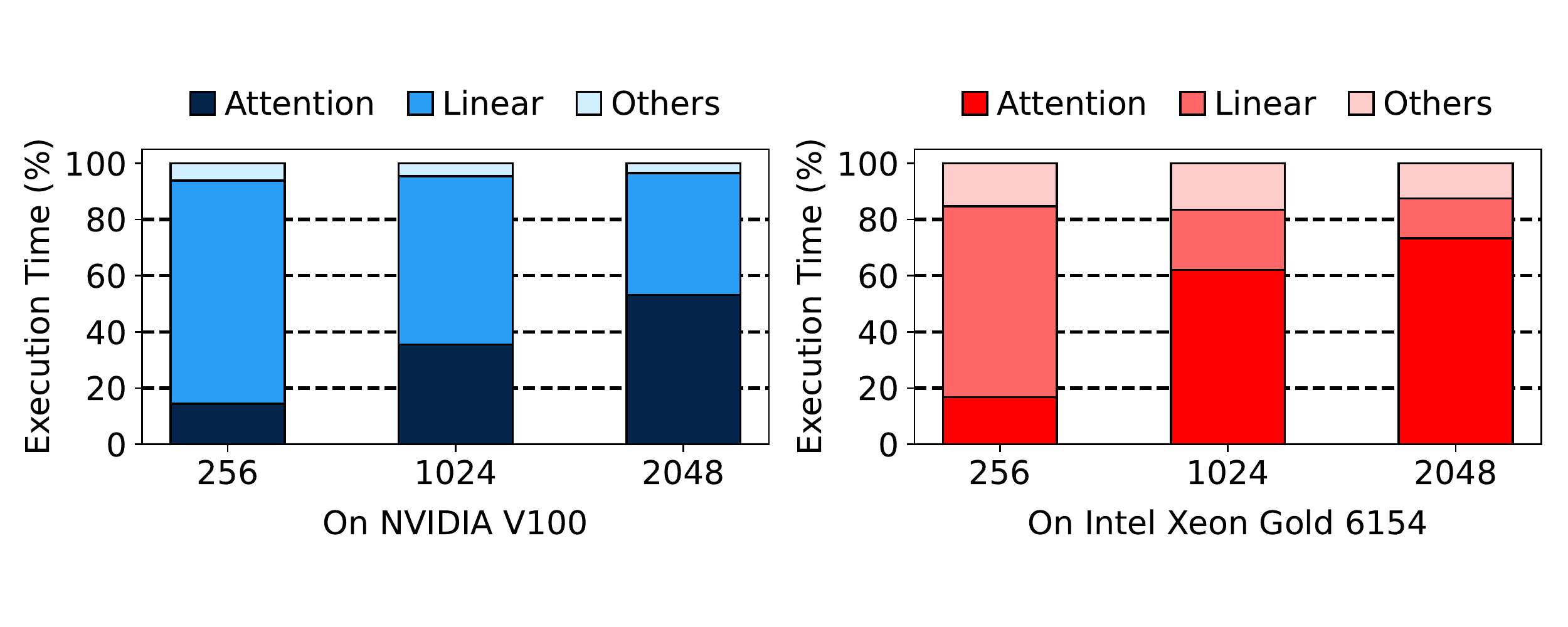}
\vspace{-4.0mm}
\caption{Execution time breakdown of \textit{Transformer} with different input lengths on GPU and CPU. }\label{fig:latency_breakdown}
\vspace{-3.0mm}
\end{figure}

The operation counts in~\figref{fig:op_count} reveal that the computation of attention-based NNs is dominated by different components when the length of input sequences changes.
To further investigate the real hardware performance of each subcomponent,
we profile the execution time of the \mbox{BERT-Large} model on the Nvidia V100 GPU and Intel Xeon Gold 6154 CPU.
The length of input sequences is set to $256$, $1024$ and $2048$ on both devices,
and the batch size for GPU and CPU is $8$ and $1$, respectively.
\figref{fig:latency_breakdown} shows the latency breakdown.
We split the latency consumption into three main subcomponents: attention layers, linear layers, and other operations, \textit{e.g.}~layer normalization, residual connections, matrix transformations and IO operations. 
Notably, on both CPU and GPU, linear layers take up a significant portion of execution time, $67.9\%$ and $79.34\%$ respectively, when the input length is small. 
As the input length becomes larger, the execution time of attention layers increases gradually and becomes dominant.
As such,
the latency is dominated by different components depending on the length of the input sequence.
According to Amdahl's law~\cite{amdahl1967validity},
to achieve high hardware performance across different input lengths,
it is necessary to optimize both attention and linear layers.

The majority of previous accelerators for attention-based NNs focused on optimizing a single component of the entire model (either attention or FFN as shown in~\tabref{tb:sota_sparse}), leading to suboptimal end-to-end performance gains. 
The execution time of these accelerators is heavily dependent on the input length which varies across different applications, reducing the scalability of these hardware designs and thus narrowing their deployability in real-world scenarios.
Naively adopting a combination of previous works on optimizing the linear layers~\cite{li2020ftrans} and attention layers~\cite{ham2020, wang2020spatten, sanger201micro, zhou2021energon, elsa2021isca, dota2022asplos}, however, would result in low hardware efficiency as they adopt different sparsity patterns.
As a result, designing an end-to-end accelerator for scalable attention-based NNs remains an open problem.
In this work, we address this challenge by adopting an algorithm and hardware co-design approach.
On the algorithmic level,
a hardware-friendly model called \textit{FABNet} is proposed,
which adopts a unified butterfly sparsity pattern to compress both attention and linear layers.
On the hardware level,
we propose an adaptable butterfly design that can be configured at runtime to accelerate different layers in \textit{FABNet} using one unified hardware engine.

%% file: 3_algorithm_opt.tex
\section{Algorithm Optimization}\label{sec:alg_opt}
\subsection{Computational Analysis of Sparsity Patterns}\label{subsec:sparsity}

\begin{figure}[t]
\centering
\includegraphics[width=0.49\textwidth]{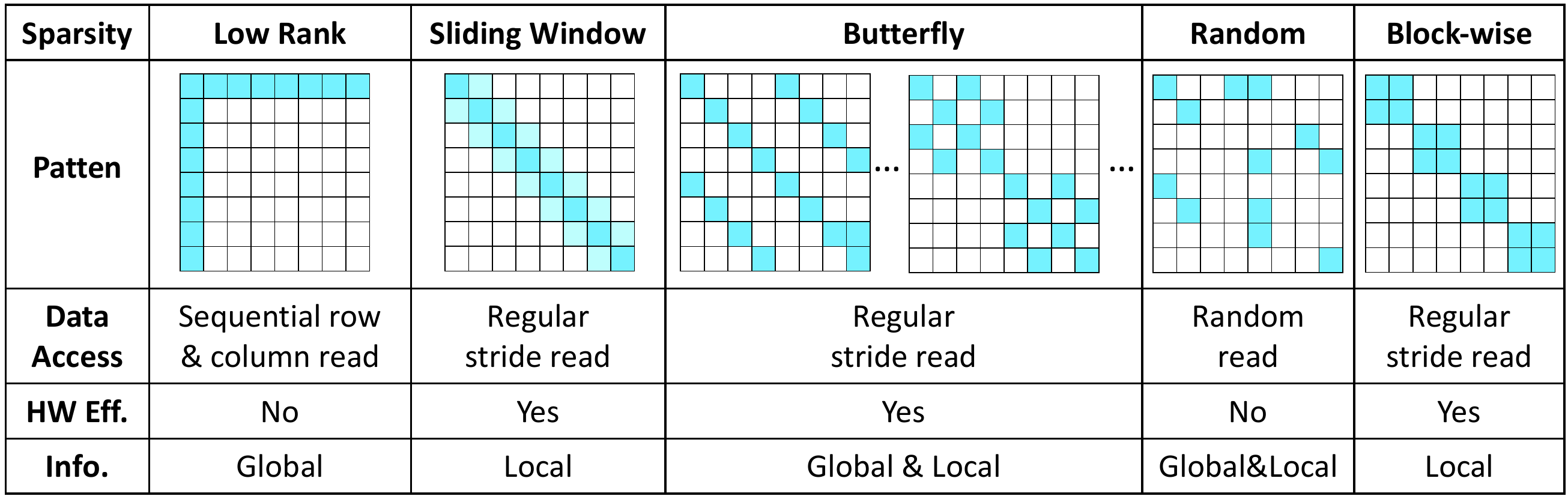}
\vspace{-3.5mm}
\caption{{Basic sparsity patterns in existing variants.}}
\vspace{2.0mm}
\label{fig:sparsity}
\end{figure}

Various pruning schemes have been proposed to reduce the computational complexity of attention-based NNs, leading to different efficient models~\cite{choromanski2020rethinking, wang2020linformer, kitaev2020reformer, tay2020sparse, beltagy2020longformer, zaheer2020big, child2019generating, chen2021pixelated, lee2021fnet, dao2020kaleidoscope, dao2022monarch}.
By analysing the computational and data access patterns of these variants,
we define five basic sparsity patterns shown in~\figref{fig:sparsity}: \textit{i)}~low rank, \textit{ii)}~sliding window, \textit{iii)}~butterfly, \textit{iv)}~random, and \textit{v)}~block-wise pattern.
As low-rank approximation of an attention matrix requires both sequential row and column reads but the data are usually only stored in either a row-major or column-major, the hardware efficiency of low-rank sparsity is inherently diminished.
Random sparsity also demonstrates low hardware efficiency due to its random read pattern.
Furthermore, we observe that the sparsity in various sparse variants can be expressed as different combinations of the basic sparsity patterns, as summarized in
\tabref{tb:sparse_att}.
As some basic sparsity patterns can only capture either long-range global or short-range local information (\figref{fig:sparsity}),
the rationale behind using multiple sparsity patterns within each variant is mainly to compensate for the underlying accuracy loss. 
For example,
Pixelfly~\cite{chen2021pixelated} introduces an additional low-rank sparsity pattern to increase the expressiveness of their flat block-wise butterfly pattern and improve accuracy.

\begin{table}[t]
\centering
\caption{Combination of sparsities in different variants.}
\vspace{2.0mm}
\label{tb:sparse_att}
\setlength\tabcolsep{1pt} 
\scalebox{0.83}{
% \begin{tabular}{L{2cm}ccC{1.5cm}C{2.5cm}}
\begin{tabular}{C{2.8 cm}|C{2.9cm}|C{0.8cm} C{0.9cm} C{1.25cm} C{1.25cm}}
\toprule
{\textbf{Model} }& { \textbf{Sparsity pattern} }& {\textbf{Att.}}& {\textbf{FFN}}& {\textbf{Unified Sparsity}} & {\textbf{Co-Design}}\\ 
\midrule
{Performer~\cite{choromanski2020rethinking}}& {Low-Rank} & \multirow{2}{*}{\ding{52}} & \multirow{2}{*}{\ding{55}} & \multirow{2}{*}{\ding{55}}& \multirow{2}{*}{\ding{55}}\\
Linformer~\cite{wang2020linformer} & (Extra kernels) & & & &\\
\midrule
\multirow{2}{*}{Reformer~\cite{kitaev2020reformer}}& {Block-wise} & \multirow{2}{*}{\ding{52}} & \multirow{2}{*}{\ding{55}} & \multirow{2}{*}{\ding{55}}& \multirow{2}{*}{\ding{55}}\\
& {(Extra kernels)} & & & &\\
\midrule
{Sparse Sinkhorn~\cite{tay2020sparse}}& {Block-wise + Random} & \ding{52} & \ding{55}& \ding{55} & \ding{55}\\
\midrule
{ Longformer~\cite{beltagy2020longformer}}& {Sliding-Window + Low-Rank} & \ding{52} & \ding{55} & \ding{55}& \ding{55}\\
\midrule
{ BigBird~\cite{zaheer2020big}}& {Random + Sliding-Window + Low-Rank} & \ding{52} & \ding{55} & \ding{55}& \ding{55}\\
\midrule
{FNet~\cite{lee2021fnet}}& {Butterfly} & \ding{52} & \ding{55} & \ding{55}& \ding{55}\\
\midrule
{Kaleidoscope~\cite{dao2020kaleidoscope}}& {Butterfly} & \ding{55} & \ding{52} & \ding{52}& \ding{55}\\
\midrule
{Sparse Trans.~\cite{child2019generating}}& {Low-Rank + Butterfly + Sliding-Window} & \ding{52} & \ding{55} & \ding{55}& \ding{55}\\
\midrule
{ Pixelfly~\cite{chen2021pixelated} Monarch~\cite{dao2022monarch}}& {Butterfly + Block-Wise + Low-Rank} & \ding{52} & \ding{52} & \ding{55}& \ding{55}\\
\midrule
{ \textbf{Our work}}& {Butterfly} & \ding{52} & \ding{52} & \ding{52}& \ding{52}\\
\bottomrule
\end{tabular}}
\vspace{-2.0mm}
\end{table}

Different sparsity patterns exhibit diverse data access patterns, which calls for custom hardware support.
However, supporting multiple sparsity patterns may complicate the hardware design.
For instance, in order to fully utilize the sparsity in the random pattern, complex dynamic controllers are required to achieve a load-balanced execution on different hardware engines~\cite{sanger201micro, geng2020awb}.
The extra overhead of such controllers may counteract the improvement brought by skipping sparse operations~\cite{jang2021sparsity}.

In this work, we aim to find a hardware-friendly sparsity pattern that: \textit{1)}~has structured data access patterns to simplify the memory design, \textit{2)}~captures both local and global range information with a single sparsity pattern, and \textit{3)}~is applicable to both the attention mechanism and FFNs to sustain its performance improvement across both long and short input sequences.
To meet these requirements, we adopt the \textit{butterfly} sparsity as a basis for constructing our efficient algorithm.

Compared to other sparsity patterns, the butterfly sparsity provides a number of favorable properties.
As shown in~\figref{fig:sparsity},
although random sparsity is able to capture both local and global information,
it has two drawbacks compared to butterfly sparsity: \textit{1)}~it requires complicated controllers with excessive hardware overhead~\cite{jang2021sparsity}, and \textit{2)}~its performance gain cannot be guaranteed as the sparsity may vary substantially among different datasets and tasks.
Compared with random sparsity,
the sliding-window pattern is more hardware-friendly.
However, 
\tabref{tb:sparse_att} shows that it often requires low-rank sparsity to compensate for the accuracy loss, as sliding-window sparsity only captures the local relationship within each window.
Moreover, although some variants adopt a single low-rank or block-wise sparsity pattern with satisfactory algorithmic performance, they require extra algorithmic operations and dedicated computational kernels during inference (\textit{e.g.}~the locality-sensitive hashing (LSH) in Reformer~\cite{kitaev2020reformer}) during inference, resulting in large hardware overhead.
In contrast,
this paper treats the butterfly sparsity as a promising method due to its regular data access pattern and the ability of capturing both global and local information.

\subsection{Unified Butterfly Pattern for Attention and Linear Layers}\label{subsec:bfly_blocks}

\begin{figure}[t]
\centering
\includegraphics[width=0.49\textwidth]{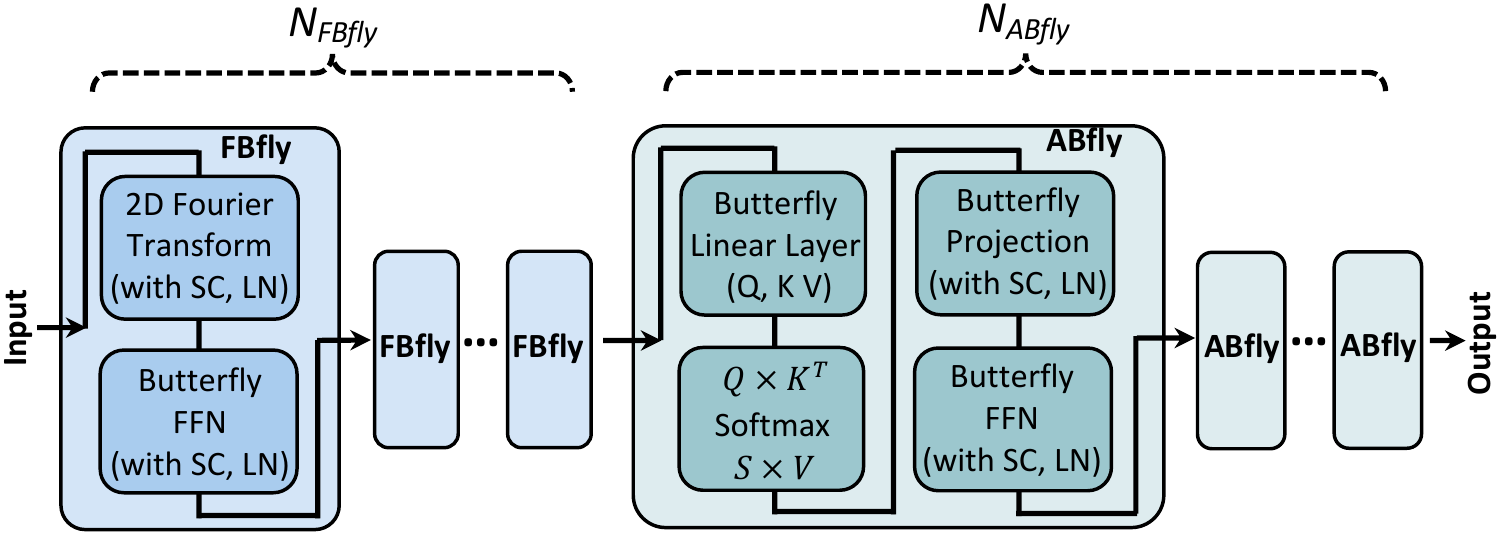}
\vspace{-1.5mm}
\caption{Network structure of \textit{FABNet}. (SC: shortcut addition, LN: layer normalization)}
\label{fig:three_bfly_fnet}
\end{figure}

\begin{figure*}[t]
\centering
\includegraphics[width=0.99\textwidth]{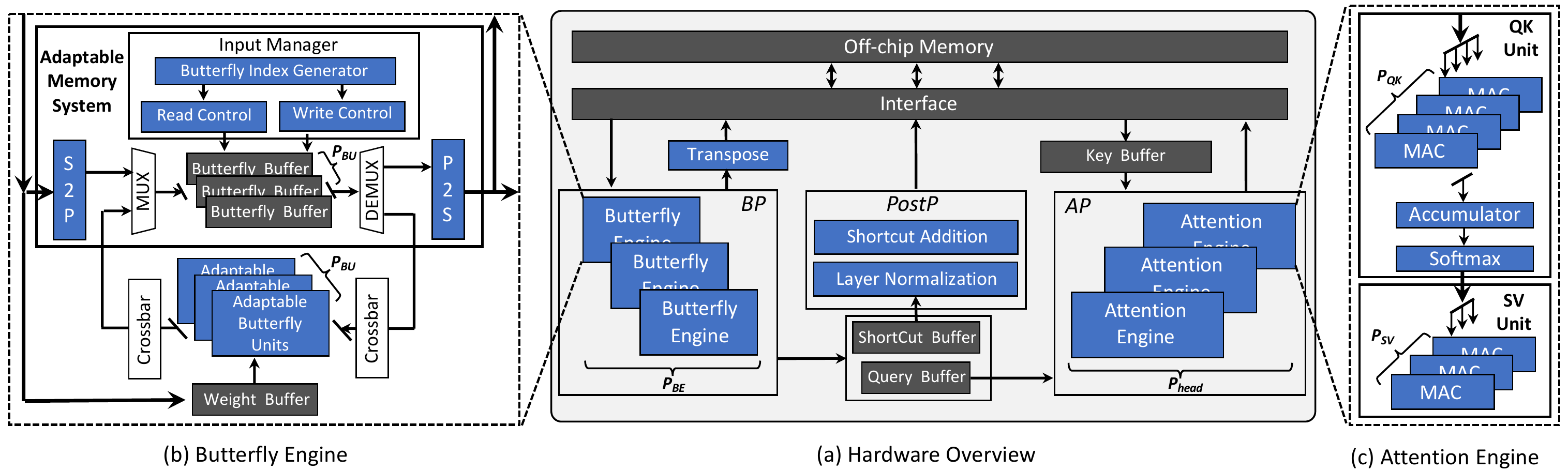}
\caption{{Hardware overview of the adaptable butterfly accelerator.}}
\label{fig:hw_overview}
\end{figure*}

\mohamed{a reviewer will ask why we keep the multi-head attention instead of approximating it with butterfly--we should address that. If we replaced all layers with butterfly/fft then we can basically build just programmable FFT hardware and we're done, which could be more attractive than leaving just one layer that still requires gemm.}
\hongxiang{This needs further software experiments to demonstrate the accuracy loss is acceptable. We don't have time for Micro, so maybe for our next paper, together with the high-radix butterfly}

The butterfly pattern has demonstrated its effectiveness and generality in approximating linear transformations~\cite{dao2020kaleidoscope}.
Furthermore, Lee-Thorp~\textit{et~al.}~\cite{lee2021fnet} have shown the potential of simplifying the computation by replacing the entire attention layer with Fourier transform, which effectively mixes tokens without explicitly approximating the attention mechanism.
% As such,
To maximize the ability to reduce the computation with acceptable algorithmic performance,
we start by proposing two basic building blocks for scalable inference: 
\textit{1)} the Attention Butterfly (\textit{ABfly}), and \textit{2)} Fourier Butterfly (\textit{FBfly}) blocks.

In the \textit{ABfly} block,
we retain the backbone of the attention module and compress all the linear layers using butterfly factorization.
Specifically, the \textit{ABfly} block starts with three butterfly linear layers to generate $\mathbf{Q}$, $\mathbf{K}$ and $\mathbf{V}$ matrices. The results are fed into a vanilla multi-head attention layer and another butterfly linear layer to obtain the relationships among different tokens.
A butterfly FFN that consists of two butterfly linear layers is placed at the end of the \textit{ABfly} block for additional processing.
To further reduce the amount of computation and number of parameters, 
we replace the attention module with a 2D Fourier transform layer, implemented using FFT, resulting in a more compute-efficient block called \textit{FBfly}.
The use of FFT effectively mixes different input tokens, which allows the following butterfly FFN to process a longer sequence.
More importantly, all computation in the FBfly block, which use the FFT's twiddle factors and the butterfly linear layers' weights, is performed using a unified butterfly pattern, resulting in higher hardware efficiency over previous works.

Although \textit{FBfly} is less compute- and memory-intensive than \textit{ABfly}, the use of the Fourier transform layer may degrade accuracy~\cite{lee2021fnet}. To preserve high accuracy, we propose a novel butterfly-based network called \textit{FABNet} that introduces a hybrid of the \textit{ABfly} and \textit{FBfly} blocks, as depicted in~\figref{fig:three_bfly_fnet}.
There are $N_{\text{FBfly}}$ \textit{FBfly} blocks at the beginning and $N_{\text{ABfly}}$ \textit{ABfly} blocks stacked on top.
With this setup, we expose both $N_{\text{FBfly}}$ and $N_{\text{ABfly}}$ as hyperparameters, enabling a trade-off between algorithmic and hardware performance.
To optimize this trade-off, we develop a co-design method (\secref{subsec:co-design}) that explores the design space of both neural architecture and hardware design.

%% file: 4_hardware_design.tex
\section{Hardware Accelerator}\label{sec:hw}
\subsection{Architecture Overview}\label{subsec:hw_overview}

\mohamed{should we add a small subsection about the attention processor? Also, do we talk later about the area dedicated for the attention engine vs. the butterfly engine? Do we configure that based on the FABNet parameters? What is the utilization of each engine?}
\hongxiang{attention processor is just two MAC unit connected together, so there is not too much to say....}

\figref{fig:hw_overview} shows the proposed hardware accelerator consisting of: a Butterfly Processor ({\textit{BP}}), an Attention Processor ({\textit{AP}}),
a Post-processing Processor ({\textit{PostP}}), the off-chip memory, and several on-chip buffers.
\textit{BP} consists of $P_{\text{BE}}$ number of Butterfly Engines ({\textit{BEs}}), which are used to accelerate the computations that involve butterfly patterns, including both FFT and butterfly linear transformations.
\textit{AP} contains $P_{\text{AE}}$ number of Attention Engines ({\textit{AEs}}),
and each {\textit{AE}} is composed of one {\textit{QK}} unit and one {\textit{SV}} unit.
The {\textit{QK}} unit is designed to implement the softmax and the matrix multiplication between queries and keys.
The {\textit{SV}} receives the outputs from the {\textit{QK}} unit and multiplies the results with value vectors to generate the final results of the attention layer.
The \textit{PostP} module is responsible for executing the layer normalization and shortcut (\textit{SC}) addition.
To ease the on-chip memory consumption,
the intermediate results between different FFT and butterfly operations are transferred back to the off-chip memory.
Although doing so increases the bandwidth requirement,
this ensures our accelerator is scalable on hardware platforms with limited on-chip memory. 
To improve the overall hardware performance,
all the on-chip buffers utilize double-buffering to overlap the data transfer with the computation.

\begin{figure}[t]
\centering
\includegraphics[width=0.49\textwidth]{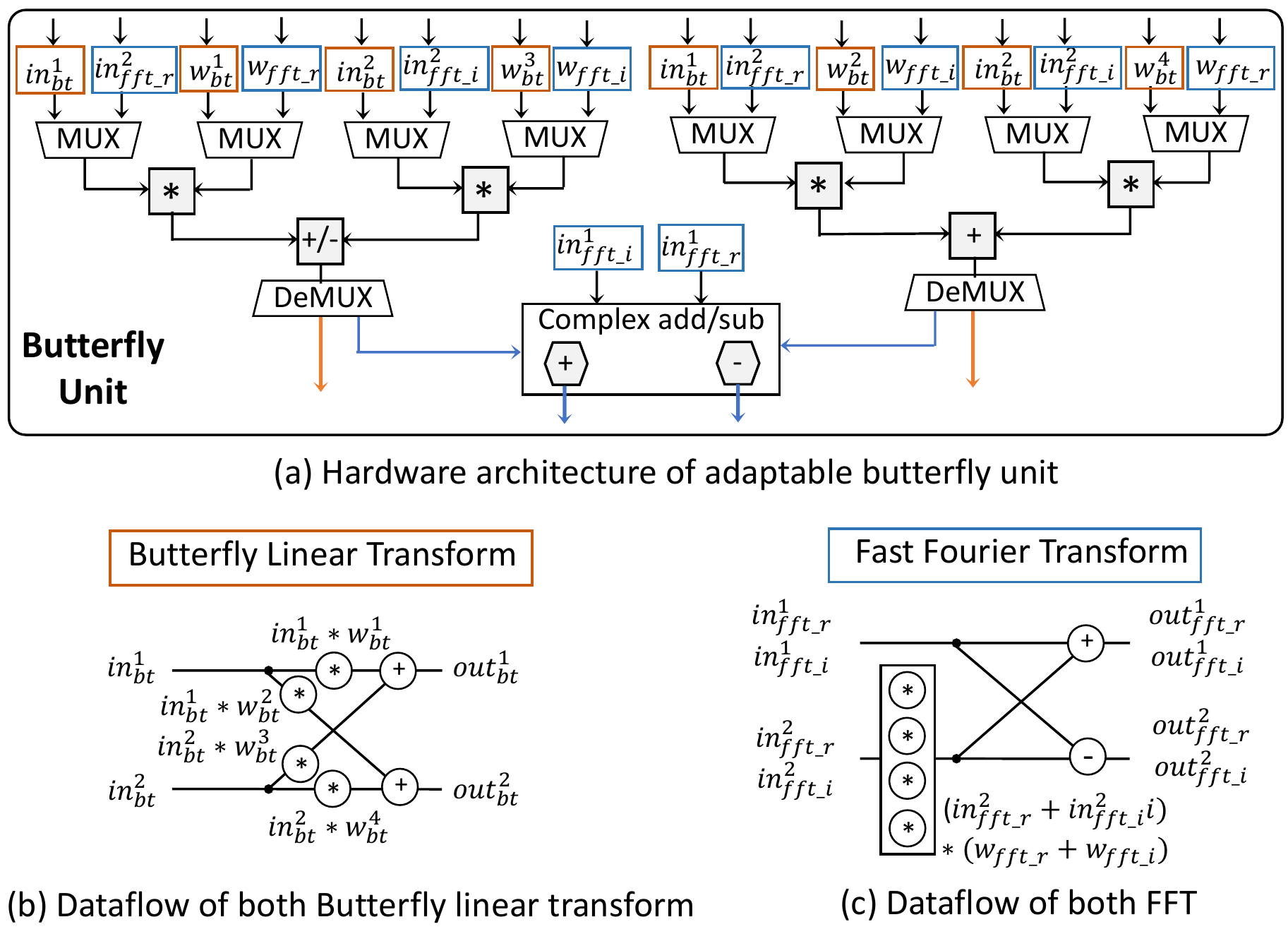}
% \vspace{-2.5mm}
\caption{Microarchitecture and dataflow of the adaptable butterfly unit.}
% \vspace{-2.5mm}
\label{fig:butterfly_unit}
\end{figure}

\subsection{Adaptable Butterfly Engine}
\figref{fig:hw_overview}b shows the hardware architecture of \textit{BE}.
Each \textit{BE} is mainly composed of a butterfly memory system and $P_{\text{BU}}$ number of adaptable Butterfly Units (\textit{BU}s).
To improve the hardware efficiency and enable the use of a single unified engine,
the \textit{BE} module is designed with a focus on adaptability. As such,
it can be configured via programmable multiplexers and de-multiplexers at runtime to either execute an FFT or a butterfly linear transformation.

\noindent
\subsubsection{Adaptable Butterfly Unit}
\figref{fig:butterfly_unit}a depicts the architecture of the proposed adaptable \textit{BU}.
Each adaptable \textit{BU} consists of four real-number multipliers and two real-number adders, followed by two complex-number adders.
The inputs and twiddle factors of both FFT and butterfly linear transformation are connected to the multipliers,
with eight multiplexers used to select the correct inputs for each operation.
Two de-multiplexers are placed after the real-number adders to control the output flow.

% There are several multiplexers and de-multiplexers in each \textit{BU} to control the inputs and outputs flows.

When performing the butterfly linear transformation (\figref{fig:butterfly_unit}b),
the twiddle factors are non-symmetric real numbers. Hence, the output of each twiddle multiply can computed as:
\begin{equation*}\label{eq:comp_butterfly}
out_{\text{bt}}^{1} = in_{\text{bt}}^{1} \cdot w^{1}_{\text{bt}} + in_{\text{bt}}^{2} \cdot w^{3}_{\text{bt}}, \quad \,
out_{\text{bt}}^{2} = in_{\text{bt}}^{1} \cdot w^{2}_{\text{bt}} + in_{\text{bt}}^{2} \cdot w^{4}_{\text{bt}}
\end{equation*}
where $in_{\text{bt}}^{1 \sim 4}$ and $w^{1 \sim 4}_{\text{bt}}$ represent the inputs and twiddle factors, respectively.
To perform the butterfly linear transformation,
four multipliers in each \textit{BE} are configured to execute the four real-number multiplications in the equation above.
The values $in_{\text{bt}}^{1 \sim 4}$ and $w^{1 \sim 4}_{\text{bt}}$ are selected via multiplexers as the operands of the multipliers.
At the same time, the results $out_{\text{bt}}^{1 \sim 1}$ generated from the real-number adders/subtractors are outputted directly from the de-multiplexers.

For FFT (\figref{fig:butterfly_unit}c),
since the twiddle factors of FFT are complex and symmetric,
it only requires one complex-number multiplication per twiddle multiplication.
Thus,
by selecting the complex inputs $in_{\text{fft}\_r}^{1 \sim 2} + in_{\text{fft}\_i}^{1 \sim 2}i$  and twiddle factor $w_{\text{fft}\_r} + w_{\text{fft}\_i}i$,
we reuse the four real-number multipliers in each \textit{BE} to perform the required complex-number multiplication.
The de-multiplexers are then configured to output the results to the complex-number adders/subtractors to get the final results $out_{\text{fft}\_r}^{1 \sim 2} + out_{\text{fft}\_i}^{1 \sim 2}i$.
The control signals for the multiplexers and de-multiplexers are set  before running each layer.
As such, the proposed adaptable \textit{BE} can be used to accelerate both FFTs and butterfly linear transformations by reusing the multipliers, adders and subtractors.
% to improve the hardware efficiency.

\noindent
\subsubsection{Butterfly Memory System}
\begin{figure}[t]
\centering
\includegraphics[width=0.49\textwidth]{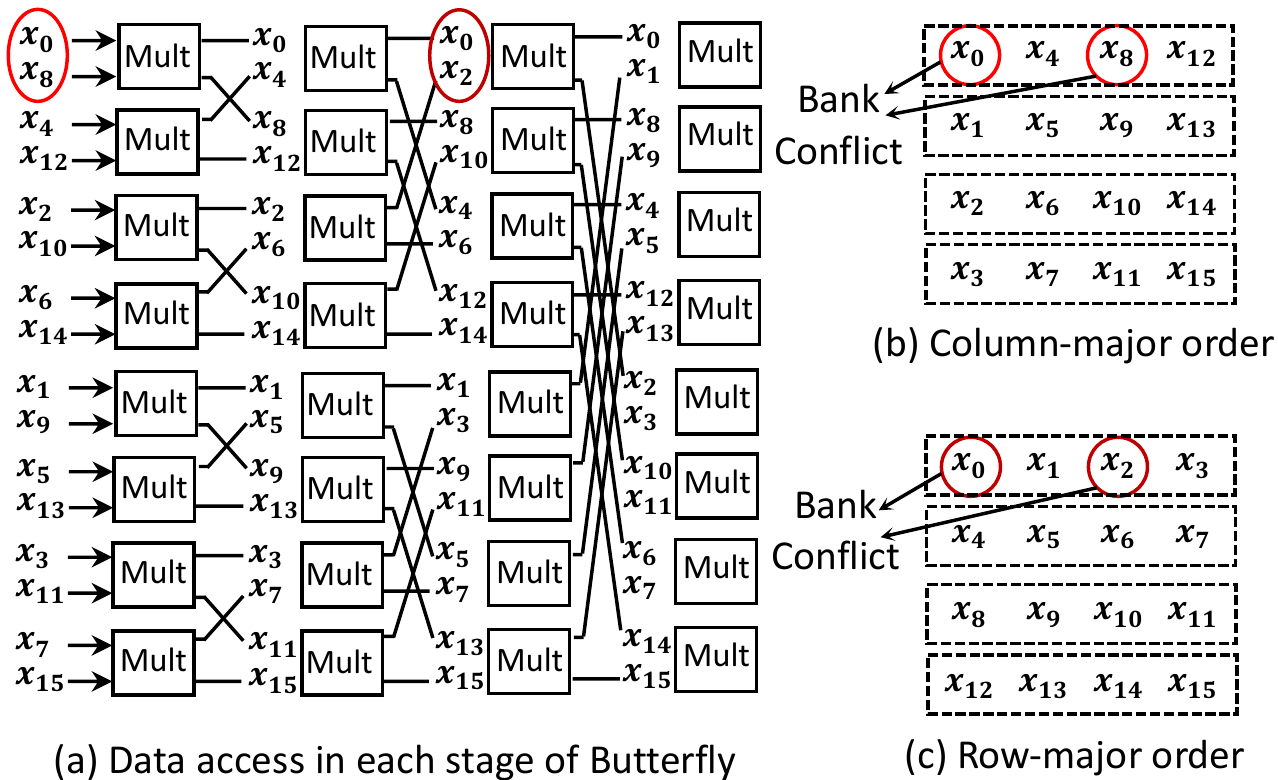}
% \vspace{-2.mm}
\caption{{Bank conflicts in column and row-major orders.}}
\label{fig:bank_conflict}
\end{figure}

Our butterfly memory system comprises an input manager, a serial-to-parallel (\textit{S2P}) module, a parallel-to-serial (\textit{P2S}) module and butterfly buffers.
As shown in~\figref{fig:bank_conflict}a,
the butterfly pattern requires different data access at different stages.
The conventional column-major or row-major order will cause bank conflicts while reading the data.
For instance,
accessing index pair $x_{0}$ and $x_{8}$ of the first stage causes a read conflict in the column-major order as shown in~\figref{fig:bank_conflict}b, in which each row represents a memory bank.
The row-major order also suffers from the same issue while reading $x_{0}$ and $x_{2}$ in the third stage.

\begin{figure}[t]
\centering
\includegraphics[width=0.49\textwidth]{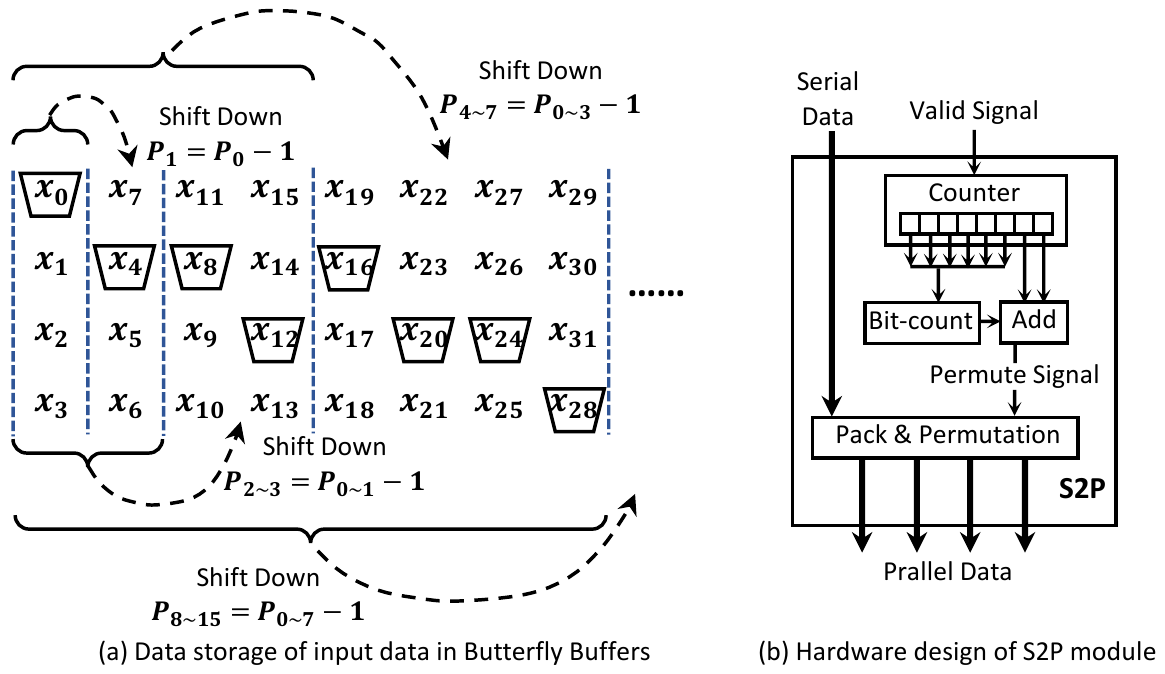}
\caption{{Data layout and hardware design of \textit{S2P}.}}
% \vspace{-2.5mm}
\label{fig:hw_s2p}
\end{figure}

To avoid such bank conflict,
we introduce a custom data layout strategy and implement it using the \textit{S2P} module shown in~\figref{fig:hw_s2p}.
We permute each column $i$ using a starting position $P_{i}$ which indicates how many rows the first element in the current column should be shifted down.
We define the starting position using the following formula:
\begin{equation*}\label{eq:permute_formula}
P_{0} = 0, \quad
P_{2^{n-1} \sim (2^{n}-1)} = P_{0 \sim (2^{n-1}-1)} - 1 \; \text{for} \; 1 \leq n \leq N
\end{equation*}
For each $2^{n-1} \sim (2^{n}-1)$ columns, the starting positions $P_{2^{n-1} \sim (2^{n}-1)}$ is obtained by shifting  $P_{0 \sim (2^{n-1}-1)}$ one position down, as shown in~\figref{fig:hw_s2p}a.
The starting positions are generated using a counter, and a bit-count and addition operations (\figref{fig:hw_s2p}b).
After packing the serial data together,
\textit{S2P} permutes them based on the starting positions.

\begin{figure}[t]
\centering
\includegraphics[width=0.46\textwidth]{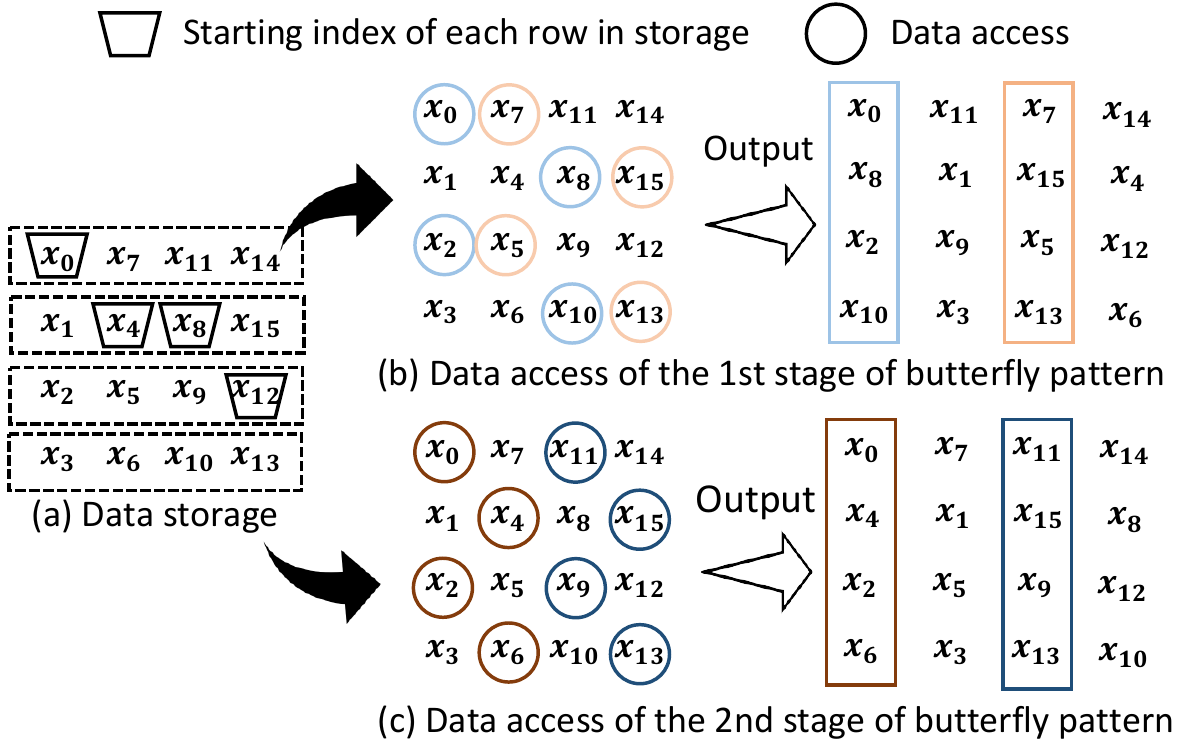}
\caption{{An example of 16-input butterfly.}}
% \vspace{-0.5cm}
\label{fig:data_bfly}
\end{figure}

\begin{figure}[b]
\centering
\includegraphics[width=0.49\textwidth]{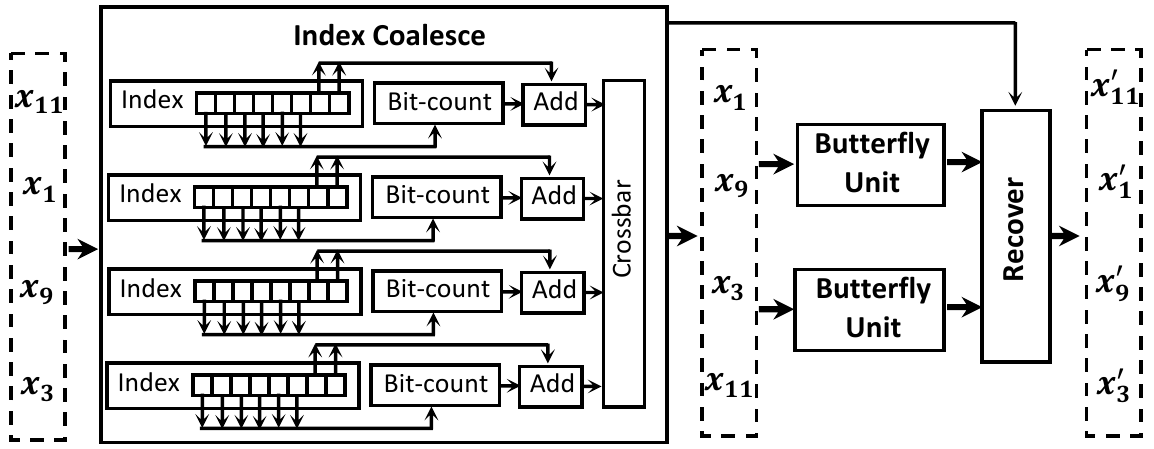}
% \vspace{-4mm}
\caption{{Hardware design of \textit{Index Coalescing} module.}}
\label{fig:hw_rotation}
\end{figure}

\figref{fig:data_bfly} presents an example with 16 inputs,
where the data required by the first and second stage of the butterfly pattern are read from the buffers without bank conflicts.
However,
as the butterfly units receive data in pairs,
an extra pairing is required after the \textit{S2P} module.
An example is the second output column $\langle x_{11}, x_{1}, x_{9}, x_{3} \rangle$ of the first stage in~\figref{fig:data_bfly}b.
To pair indices, we design an index coalescing module before the butterfly units (\figref{fig:hw_rotation}).
Based on the index of each input,
a bit-count and addition operation is used to calculate the corresponding shift position.
Then, a crossbar coalesces the index pairs based on the indices and shift positions.
To ensure the outputs generated from the butterfly units preserve the original order, a recover module is used before the data is written back.

%% file: 5_hardware_opt.tex
\section{Optimizations and Co-Design}\label{sec:hw_opt}

\noindent
\subsection{Memory Sharing in Butterfly Buffers}
We employ butterfly buffers to allow the overlap between data transfer and computation.
To reduce the memory consumption and improve the hardware efficiency,
the butterfly buffers are shared between both FFT and butterfly linear transformation.
Nonetheless, as the data width of FFT is twice that of the butterfly linear transformation,
different address mapping and overlapping strategies are required.

\begin{figure}[t]
\centering
\includegraphics[width=0.39\textwidth]{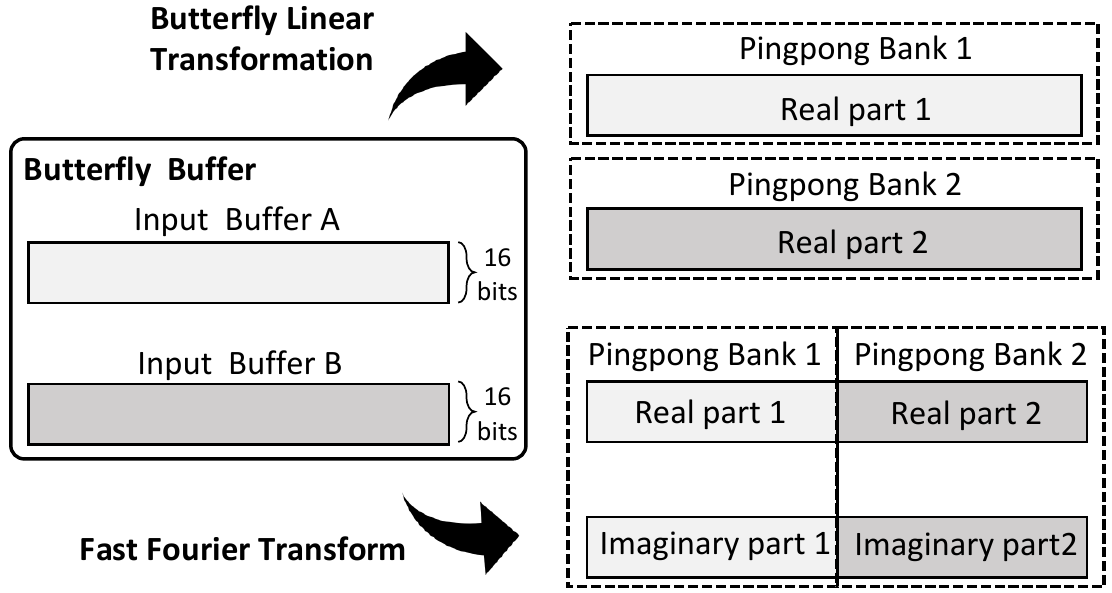}
\caption{{Different address mapping strategies.}}
\label{fig:butterfly_memory}
\end{figure}

\begin{figure}[t]
\centering
\includegraphics[width=0.49\textwidth]{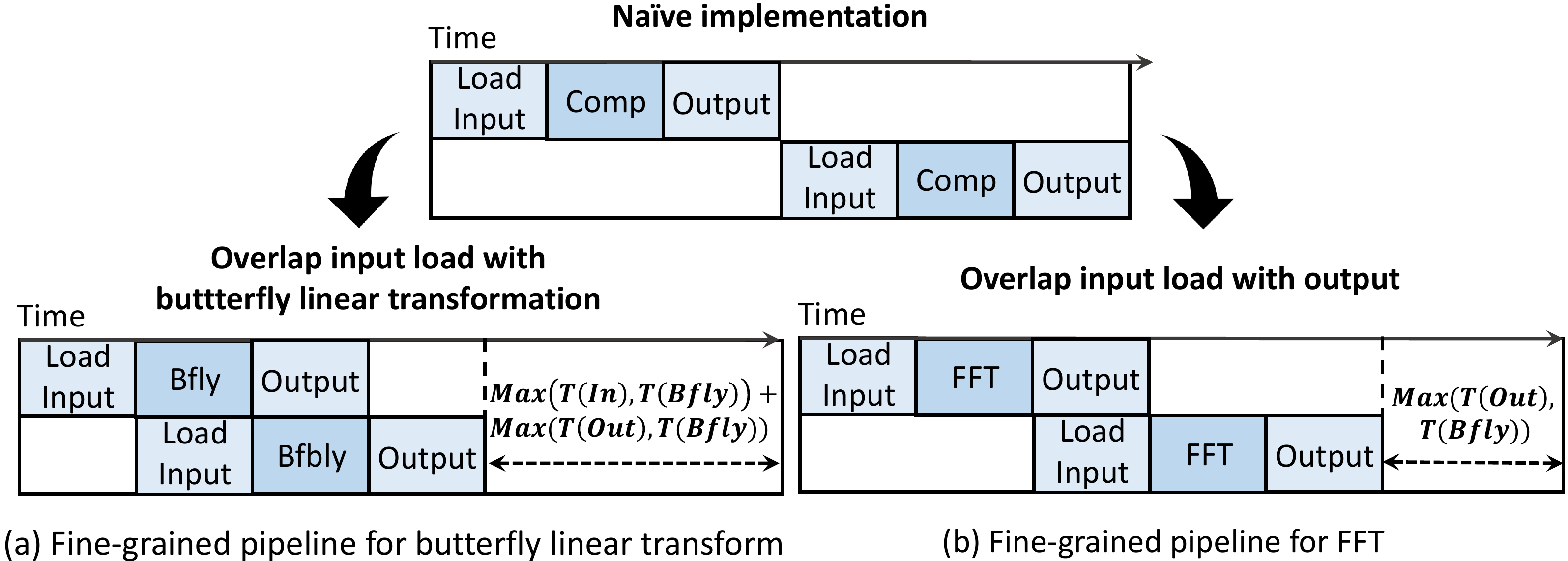}
\vspace{-2mm}
\caption{{Different overlapping strategies.}}
\vspace{-2.0mm}
\label{fig:pipeline}
\end{figure}

\figref{fig:butterfly_memory} shows the proposed address mapping strategies for butterfly linear transformation and FFT.
Assuming the bitwidth of real numbers is 16 bits, each input buffer is 16-bit wide.
While processing butterfly linear transformations, 
input buffers \textit{A} and \textit{B} are used as two independent ping-pong banks with separate read and write ports (top right in \figref{fig:butterfly_memory}).
In this manner,
when input buffer \textit{A} is used for computation,
buffer \textit{B} can start the input data transfer for the next batch, leading to the overlapping strategy shown in~\figref{fig:pipeline}a.
While processing FFT, since the data include both real and imaginary parts which require 32-bit read and write ports,
we concatenate the lower parts of input buffer A and B as the first ping-pong bank for the storage of complex numbers.  
To improve the hardware efficiency, we further reuse the higher parts of both buffers as the second ping-pong bank.   
As the computation requires both read and write accesses, we adopt a different overlapping strategy that pipelines the output data transfer only with the input data load of the next batch (\figref{fig:pipeline}b).  
By employing different address mapping and overlapping strategies for FFT and butterfly linear transformation, we maximise the hardware efficiency and performance.

\subsection{Fine-Grained Pipelining between BP and AP}
While executing the \textit{ABfly} block,
 \textit{BP} and \textit{AP} are in use, performing butterfly linear transformation and attention matrix multiplication, respectively.
To further improve performance when executing the \textit{ABfly} block,
we employ fine-grained pipelining between \textit{BP} and \textit{AP}.

\begin{figure}[t]
\centering
\includegraphics[width=0.49\textwidth]{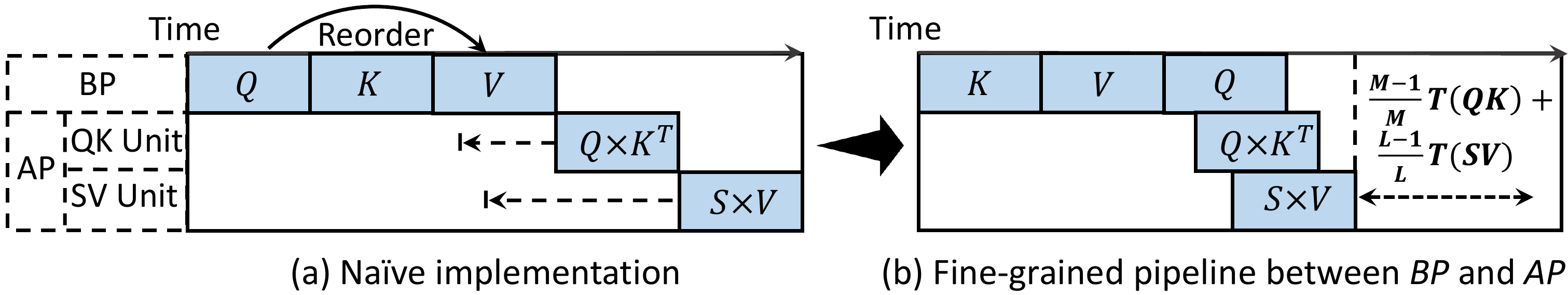}
\vspace{-3.5mm}
\caption{{Fine-grained pipelining between \textit{BP} and \textit{AP}.}}
\label{fig:pipeline_ap_bp}
\end{figure}

\figref{fig:pipeline_ap_bp} illustrates the dataflow of \textit{BP} and \textit{AP}.
In the naive implementation,
the key ($\mathbf{K}$), value ($\mathbf{V}$) and query ($\mathbf{Q}$) matrices are generated sequentially from \textit{BP}.
After $\mathbf{Q}$, $\mathbf{K}$ and $\mathbf{V}$ are computed,
\textit{AP} starts the computation of \mbox{$\mathbf{Q} \times \mathbf{K}^{T}$} and \mbox{$\mathbf{S} \times \mathbf{V}$}.
To optimize this process,
we reorder the execution sequence of linear layers such that \textit{BP} computes $\mathbf{K}$ and $\mathbf{V}$ at the beginning~(\figref{fig:pipeline_ap_bp}b).
As $\mathbf{Q} \times \mathbf{K}^{T}$ can be decomposed into multiple vector matrix multiplications that multiply different rows of $\mathbf{Q}$ with the entire matrix $\mathbf{K}^{T}$,
we can actually start the computation of $\mathbf{Q} \times \mathbf{K}^{T}$ once the first few rows of $\mathbf{Q}$ become available.
As such,
the $\mathbf{Q} \times \mathbf{K}^{T}$ in \textit{AP} can be pipelined with the computation of $\mathbf{Q}$ in \textit{BP}.
At the same time,
since $\mathbf{S}$ is generated from the \textit{QK} unit in a  row-by-row fashion,
we can further pipeline the $\mathbf{Q} \times \mathbf{K}^{T}$ with $\mathbf{S} \times \mathbf{V}$, as the computation of \mbox{$\mathbf{S} \times \mathbf{V}$} can start once the first few rows of $\mathbf{S}$ are generated from the \textit{QK} unit. Assuming there are $M$ and $L$ rows in $\mathbf{Q}$ and $\mathbf{K}$ matrices, it takes $\frac{T(QK)}{M}$ and $\frac{T(SV)}{L}$ to compute one row in the \textit{SV} and \textit{QK} units, respectively. As such, the total latency reduction achieved is $\frac{M-1}{M}T(QK) + \frac{L-1}{L}T(SV)$ compared to the unoptimized non-pipelined implementation.

\begin{figure}[t]
\centering
\includegraphics[width=0.49\textwidth]{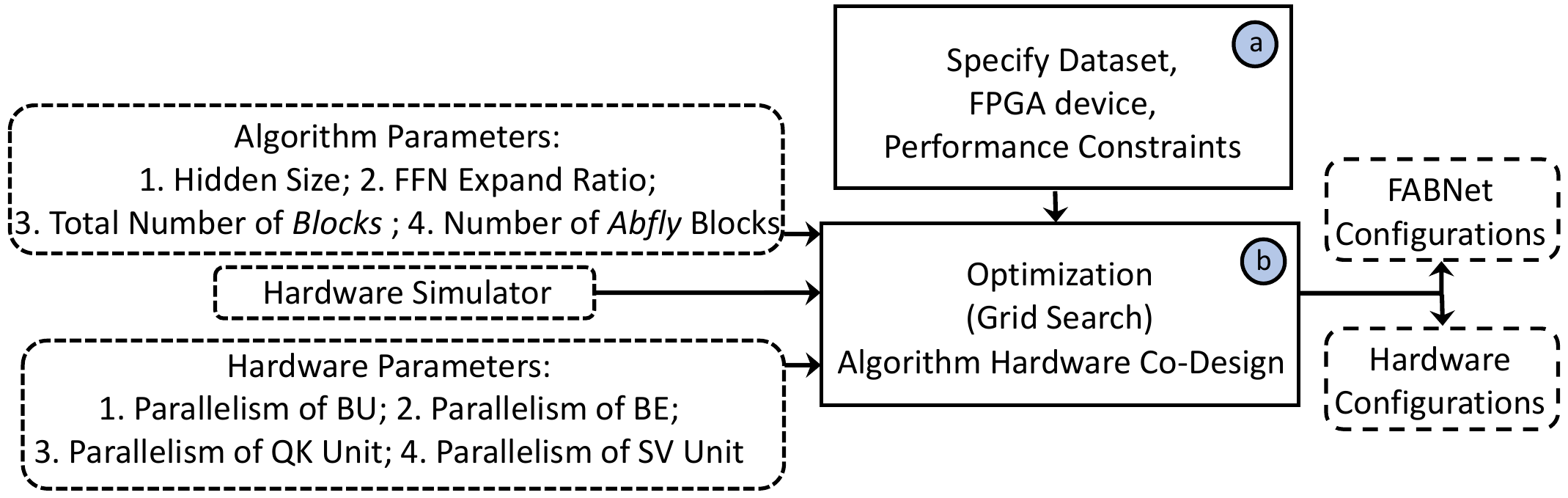}
\vspace{-3.5mm}
\caption{{Flow of the algorithm-hardware co-design process.}}
\vspace{-1.0mm}
\label{fig:co-design}
\end{figure}

\subsection{Algorithm and Hardware Co-Design}\label{subsec:co-design}

The overall design space of our end-to-end system is formed by \textit{FABNet}'s hyperparameters and the butterfly accelerator's hardware parameters.
Specifically, the joint design space consists of:
\textit{1)}~the algorithm parameters, \textit{i.e.}~the hidden size ($D_{\text{hid}}$), the expand ratio of FFN ($R_{\text{ffn}}$), the total number of blocks ($N_{\text{total}}$) and the number of \textit{ABfly} blocks  ($N_{\text{ABfly}}$) in \textit{FABNet}, and
\textit{2)}~the hardware parameters, \textit{i.e.}~the parallelism of \textit{BU} ($P_{\text{bu}}$) and \textit{BE} ($P_{\text{be}}$) in \textit{BP}, and the parallelism of the \textit{QK} ($P_{\text{qk}}$) and \textit{SV} ($P_{\text{sv}}$) units in \textit{AP}. 

To assess the trade-off provided by each design point, we need to evaluate its algorithmic performance (\textit{e.g.}~an accuracy metric), its latency and its resource consumption.
During search, the algorithmic performance is obtained by training and evaluating \textit{FABNet}, while the latency is estimated by utilizing a custom simulator built for our butterfly accelerator.
To verify whether the design can be accommodated by the target FPGA device, we developed an analytical model to estimate the consumption of DSP blocks and on-chip memory (BRAMs).
As DSPs are mainly consumed by the multipliers in \textit{AP} and \textit{BP},
we formulate its resource usage as:
\begin{align*}
    DSP_{\text{usage}} = P_{\text{be}} \times P_{\text{bu}} \times 4  +  P_{\text{head}} \times (P_{\text{qk}} + P_{\text{sv}})
\end{align*}
where the value of $4$ reflects the number of multipliers in each \textit{BU}.
The consumption of BRAM is mainly occupied by the shortcut buffer, query buffer, key buffer and different buffers in \textit{BU} including butterfly buffer and weight buffers,
which can be formulated as:
\begin{align*}\footnotesize
    BRAM_{\text{usage}} & = (BRAM_{\text{bfly}} + BRAM_{\text{weight}}) \times P_{\text{be}} \\ & + BRAM_{\text{key}}  + BRAM_{\text{sc}} + BRAM_{\text{query}}
\end{align*}
The proposed analytical resource model is only used during the design space exploration stage. 
At the end of the co-design process, the final performance is obtained by running synthesis and place-\&-route on our design with the optimized configurations.

\figref{fig:co-design} illustrates the proposed co-design approach.
Given a target dataset, FPGA device and both algorithmic and hardware performance constraints, we employ exhaustive grid search to traverse the joint design space and find the Pareto-optimal set of algorithmic and hardware parameters.
Each individual design point corresponds to a different compression ratio of \textit{FABNet} and level of parallelism of the butterfly accelerator, and provides different accuracy, latency and resource consumption.
The final output is the Pareto front of parameters for both \textit{FABNet} and our butterfly accelerator that satisfies a given set of constraints.
\vspace{3mm}

%% file: 6_evaluation.tex
\section{Evaluation}\label{sec:evaluation}
\noindent
\subsection{Experimental Setup}\label{subsec:eval_setup}
\noindent
\textbf{Benchmarks.}
To evaluate the algorithmic and hardware performance of our approach on workloads with long sequences, we choose five tasks from Long-Range-Arena~\cite{tay2020long},
including hierarchical data classification (\textit{ListOPs}), byte-level text classification (\textit{Text}), byte-level document retrieval (\textit{Retrieval}), image classification for sequences of pixels (\textit{Image}), classification of long-range spatial dependency (\textit{Pathfinder}).
The input sequences of these datasets range from $1024$ to $4096$.

\noindent
\textbf{Software Implementation.}
% \subsubsection{Software Implementation}
We implement the vanilla \textit{Transformer}~\cite{devlin2018bert}, \textit{FNet}~\cite{lee2021fnet} and our \textit{FABNet} models using \mbox{PyTorch}~{(v$1.10$)}~\cite{pytorch}.
The pretrained models are obtained from Huggingface~{$4.16$}~\cite{wolf2019huggingface}.
The batch size is $256$ for both \textit{Image} and \textit{Pathfinder} tasks, and $32$ for the rest of datasets during training.
The learning rate is set to $0.0001$, except for the \textit{Image} and \textit{Pathfinder} tasks where we use $0.01$ and $0.0005$ respectively.
Multiple Nvidia A100 and V100 GPUs are used for training.
To use FFT cores on Nvidia GPUs, the PyTorch API ``\textit{rfft2}" is used to implement the FFT operation required in both \textit{FNet} and \textit{FABNet}. 
The high-performance CUDA implementation~\cite{dao2020kaleidoscope} of butterfly linear transformation is adopted to accelerate both GPU training and inference.
We define two models with different default settings: \textit{FABNet-Base} ($D_{\text{hid}}=768$, $R_{\text{ffn}}=4$, $N_{\text{total}}=12$, $N_{\text{ABfly}}=0$) and \textit{FABNet-Large} ($D_{\text{hid}}=1024$, $R_{\text{ffn}}=4$, $N_{\text{total}}=24$, $N_{\text{ABfly}}=0$).

\noindent
\textbf{Hardware Implementation.}
We implement our hardware accelerators using Verilog.
To evaluate performance in different scenarios,
two Xilinx FPGA boards are used in our experiments: VCU128 for cloud/server scenarios and Zynq 7045 for edge/mobile settings.
Xilinx Vivado 2019.1 is used for synthesis and implementation.
While the maximum clock frequencies of our designs depend on the particular FPGA board and resource consumption, all the FPGA designs are clocked at 200 MHz which is below the maximum.
We obtain power consumption values using the Xilinx Power Estimator (XPE) tool and develop a cycle-accurate performance model to evaluate the speed performance, which is cross-validated\footnote{\small We cross-validate the functionality and correctness of our RTL design with the ground-truth results generated from PyTorch. Please refer to Appendix~\ref{sec:ae} for details.} with our RTL simulation results generated by Vivado.
The memory accesses to external memory are also considered. 
We use 16-bit half-precision floating-point in our hardware designs.
We deploy four multipliers in each \textit{BU}.
As the hidden dimension $D_{\text{hid}}$ is usually at most $1024$,
we set the depth of butterfly, query and key buffers as $1024$.
Finally, the size of shortcut buffers is the same as butterfly buffers.

\subsection{Algorithmic Performance}\label{subsec:eval_algo}

\begin{figure}[t]
\centering
\includegraphics[width=0.49\textwidth]{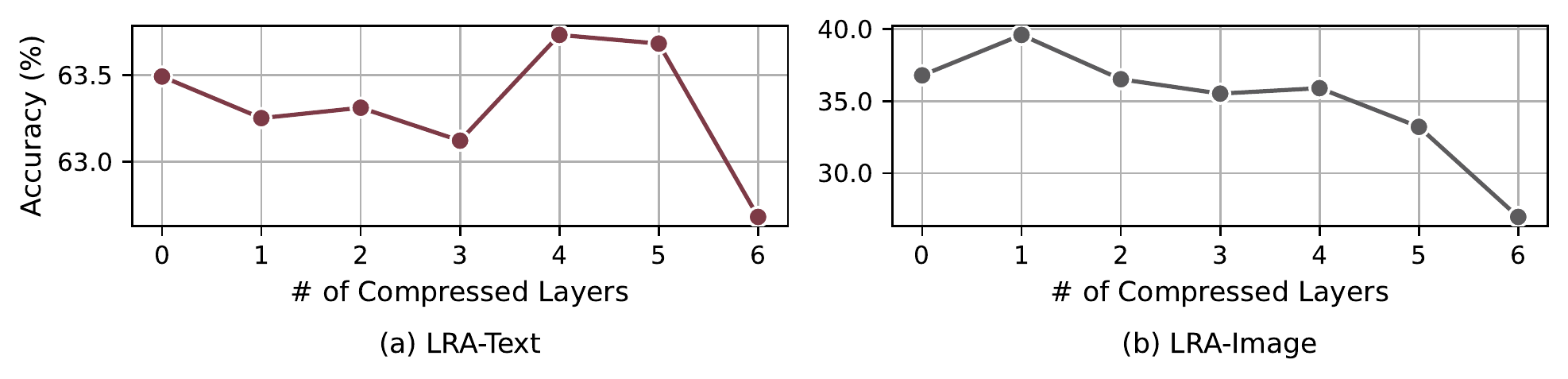}
\vspace{-5mm}
\caption{Accuracy with different number of compressed layers.}
% \vspace{3mm}
\label{fig:acc_explore}
\end{figure}

The \textit{FBfly} introduced in~\secref{subsec:bfly_blocks} is an efficient alternative to the vanilla attention block.
To evaluate its algorithmic impact on end-to-end models,
we take a six-layer \textit{Transformer} as an example\footnote{\small Other models, such as GPT and BERT, actually follow the same network architecture of \textit{Transformer} with the encoder or decoder kept. To eliminate the effect of different training strategies and evaluate the quality of the architecture, we choose the vanilla \textit{Transformer} for demonstration.} and compress it with different numbers of \textit{FBfly} blocks, starting from the last block to the first block. %ranging from $0$ to $6$.
\figref{fig:acc_explore} shows the accuracy results on LRA-\textit{Text} and LRA-\textit{Image}.
Although the accuracy fluctuates with different numbers of compressed layers,
\textit{FBfly} shows higher accuracy than the non-compressed \textit{Transformer} with $4$ and $1$ compressed layers on LRA-\textit{Text} and LRA-\textit{Image}, respectively, demonstrating the improved algorithmic performance of our approach on end-to-end models.

\begin{figure}[t]
\centering
\includegraphics[width=0.49\textwidth]{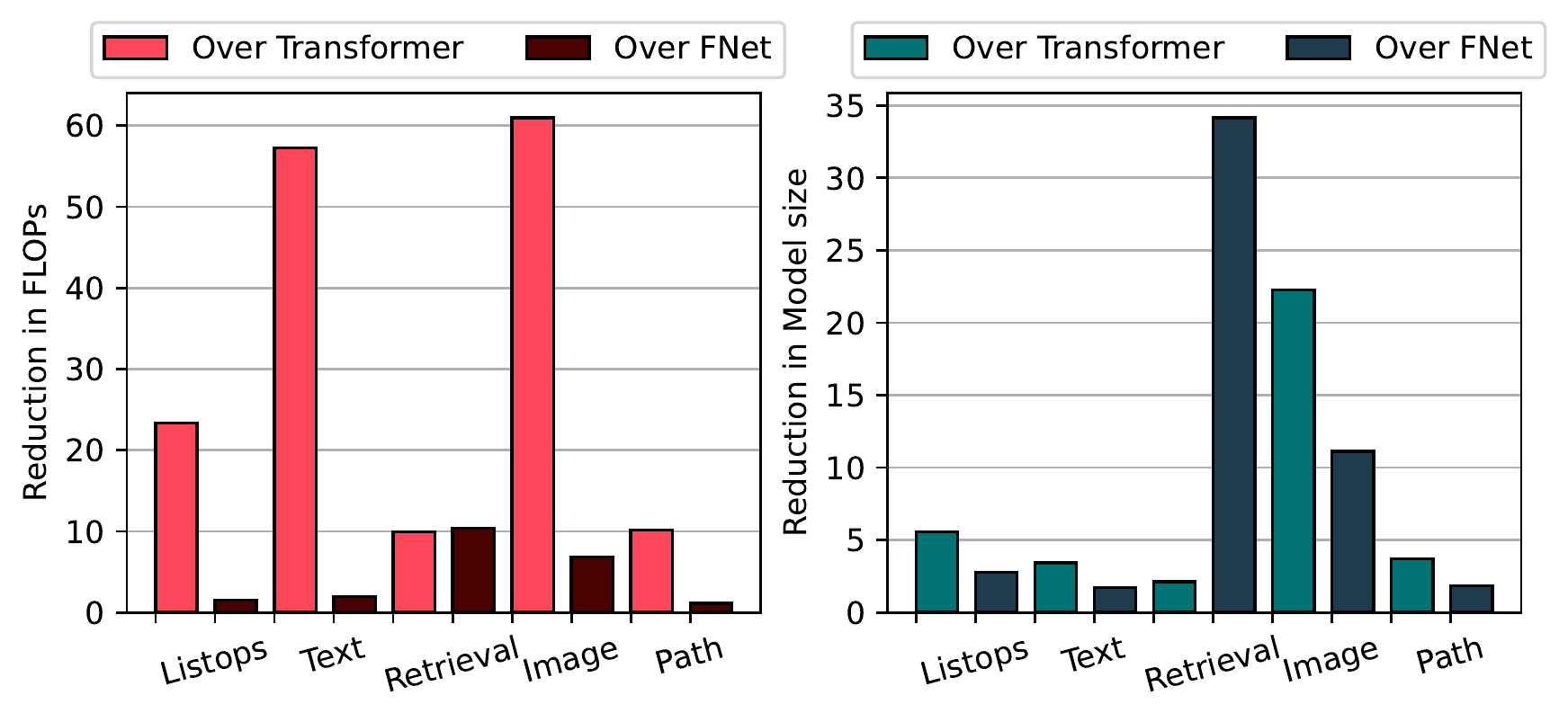}
\vspace{-6mm}
\caption{{Reduction in FLOPs and model sizes.}}
\vspace{3mm}
\label{fig:compression_rate}
\end{figure}

To obtain the best possible algorithmic performance of each model, we use the optimized configuration specified in~\cite{xiong2021nystromformer} for both vanilla \textit{Transformer} and \textit{FNet}\footnote{\small As the vanilla \textit{FNet} on \textit{Retrieval} task suffers significant accuracy loss, we increase its hidden size to $1024$.}.
We perform a simple grid search to optimize the hyperparameters of our \textit{FABNet}.
\tabref{tb:acc_lra} presents the optimized accuracy of different models.
\textit{FABNet} achieves higher accuracy than both \textit{Transformer} and \textit{FNet} on three out of five tasks, including \textit{ListOPs}, \textit{Retrieval} and \textit{Image}. 
On average, \textit{FABNet} achieves the same accuracy as \textit{Transformer}.
To investigate the efficiency of \textit{FABNet},
\figref{fig:compression_rate} shows the compression rate of our optimized \textit{FABNet} over the vanilla \textit{Transformer} and \textit{FNet} in terms of floating-point operations (FLOPs) and model size (number of parameters).
Compared with the vanilla \textit{Transformer}, \textit{FABNet} achieves around $10 \sim 60\times$ reduction in FLOPs and $2 \sim 22\times$ reduction in model size, depending on the target task.
Furthermore, compared with \textit{FNet}, \textit{FABNet} reduces FLOPs by $2 \sim 10\times$ and model size by $2 \sim 32\times$. 

\begin{figure}[t]
\centering
\includegraphics[width=0.49\textwidth]{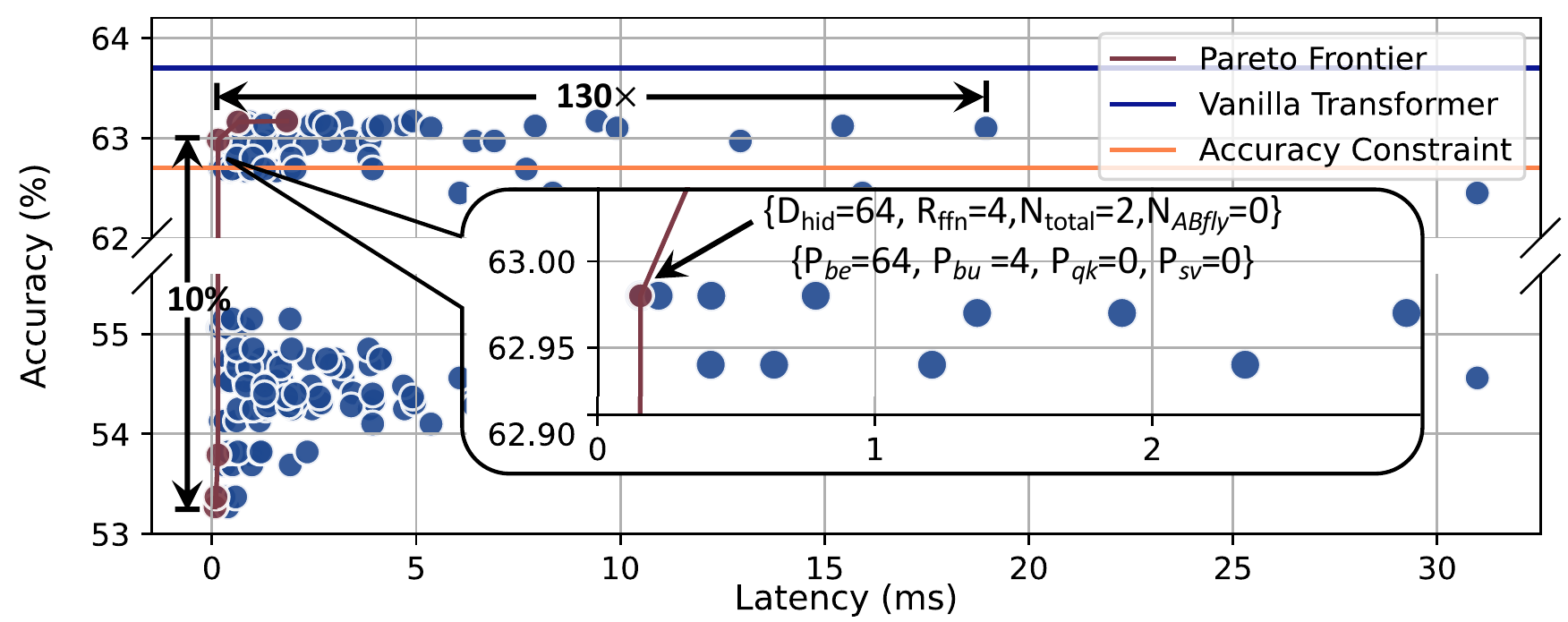}
\vspace{-5mm}
\caption{Co-design on LRA-\textit{Text} dataset.}
% \vspace{-2mm}
\label{fig:codesign_text}
\end{figure}

\begin{figure}[t]
\centering
\includegraphics[width=0.49\textwidth]{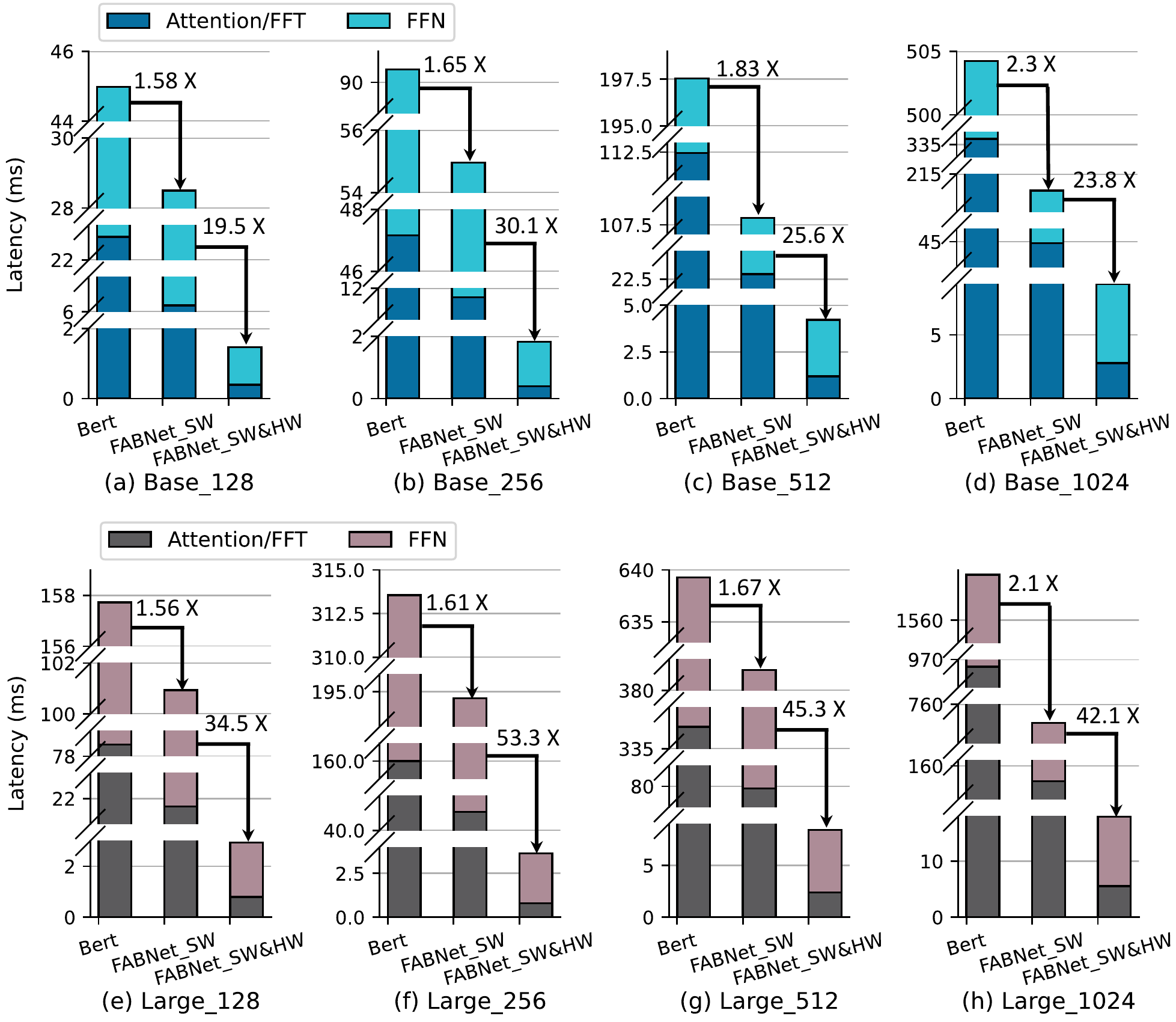}
% \vspace{-1mm}
\caption{Speedup breakdown of algorithm and hardware optimizations.}
\label{fig:breakdown_baseline}
% \vspace{-2mm}
\end{figure}

\begin{figure*}[htb]
\centering
\includegraphics[width=0.99\textwidth]{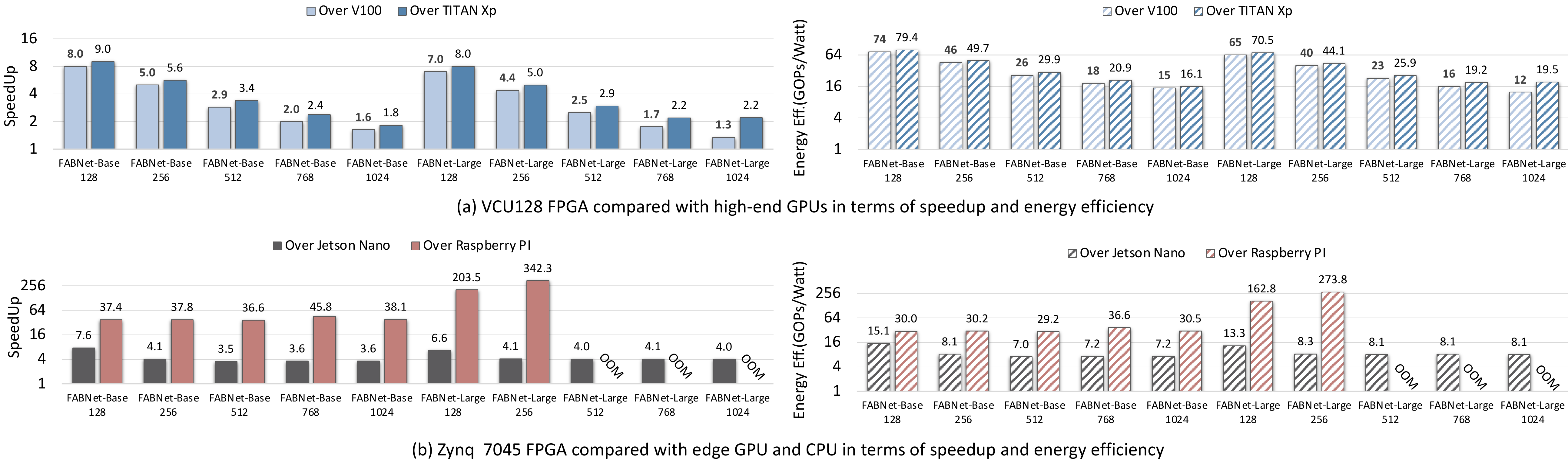}
% \vspace{-1mm}
\caption{{Performance comparison against (a) high-end GPUs, and (b) edge GPU and CPU.}}
% \vspace{-2.5mm}
\label{fig:compare_cpu_gpu}
\end{figure*}

\subsection{Effectiveness of Co-design}

We evaluate the effectiveness of our co-design approach in finding the optimal algorithm and hardware designs.
For demonstration,
we use LRA-\textit{Text} as the target dataset and VCU128 FPGA as the target device.
We select $D_{\text{hid}}$, $R_{\text{ffn}}$, $N_{\text{ABfly}}$ and $N_{\text{total}}$ from \{$64$, $128$, $256$, $512$, $1024$\}, \{$1$, $2$, $4$\}, \{$0$, $1$\} and \{$1$, $2$\} respectively.
Parameters for hardware parallelism ($P_{\text{be}}$, $P_{\text{bu}}$, $P_{\text{qk}}$ and $P_{\text{sv}}$) are chosen from \{$0$, $4$, $8$, $16$, $32$, $64$, $128$\}.
\figref{fig:codesign_text} shows the points in the accuracy-latency design space.
The orange line represents the accuracy loss, which is constrained to be less than $1$\% compared with the vanilla \textit{Transformer}.
The Pareto front is indicated by the brown line
and the other blue points represent designs with less optimized software-related hyperparameters (\figref{fig:acc_explore}) or hardware design parameters.
Among the design points that satisfy the accuracy constraint,
we choose the point with the lowest latency in the Pareto front as our point of comparison.
Within our design space, the selected point is up to $10$\% more accurate than the points in the same latency range and up to $130\times$ faster than points in the same accuracy range, underlining the advantages of our co-design approach.
The runtime of the co-design process is around $10$ hours on our GPU server.
To get the configurations for the rest of the datasets in LRA,
we constrain the overall accuracy loss to be less than $0.5$\% compared to the vanilla \textit{Transformer}.
The final models and designs are chosen as the configurations with the highest hardware performance $\left(\left<P_{\text{be}},P_{\text{bu}},P_{\text{qk}},P_{\text{sv}}\right>=\left<64, 4, 0, 0\right>\right)$
% $=64$, $P_{\text{bu}}=4$, $P_{\text{qk}}=0$, $P_{\text{sv}}=0$\}) 
without violating the accuracy constraints.
Unless mentioned otherwise,
the remaining the sections report the algorithmic and hardware performance using these optimized configurations.

\begin{table}[t]
\centering
\caption{Accuracy of different models on LRA.}
\vspace{0.2cm}
\label{tb:acc_lra}
\setlength\tabcolsep{1pt} 
\scalebox{0.78}{
% \begin{tabular}{L{2cm}ccC{1.5cm}C{2.5cm}}
\begin{tabular}{C{2.2 cm}| C{1.3cm} C{1.3cm} C{1.5cm} C{1.3cm} C{1.6cm} | C{1.cm}}
\toprule
{ }& {\textbf{ListOps}}& {\textbf{Text}}& {\textbf{Retrieval}}& {\textbf{Image}}& {\textbf{Pathfinder}} & {\textbf{Avg.}}\\ 
\midrule
{ \textbf{Vanilla \textit{Transformer}}}& {0.373} & {\textbf{0.637}} & {0.783} & {0.379} & {\textbf{0.709}} & {\textbf{0.576}}\\
\midrule
{ \textbf{Vanilla \textit{FNet}}}& {0.365} & {0.630} & {0.779} & {0.288} & {0.66} & {0.544}\\
\midrule
{ \textbf{\textit{FABNet}}}& {\textbf{0.374}} & {0.626} & {\textbf{0.801}} & {\textbf{0.398}} & {0.679} & {\textbf{0.576}}\\
\bottomrule
\end{tabular}}
\end{table}

\subsection{Comparison with Baseline Design}\label{subsec:eval_baseline}

To evaluate the speedup brought by our algorithm (\textit{FABNet}) and hardware (butterfly accelerator),
we use a baseline design for comparison~\cite{devlin2018bert}.
The baseline hardware is designed with multiple multiply-accumulate (\textit{MAC}) units to accelerate the linear transform and the matrix multiplications between query, key and value vectors.
Each \textit{MAC} is composed of a multiplier array followed by an adder tree.
The fine-grained intra- and inter-layer pipeline techniques~\cite{song2019hypar, alwani2016fused} are used to optimize the hardware performance.
We allocate the parallelism of each \textit{MAC} unit according to its workload in order to achieve load-balanced execution between different pipeline stages.
For a fair comparison,
we implement both baseline and butterfly accelerators on a VCU128 FPGA using $2048$ multipliers.
The high bandwidth memory (HBM) is used as the external memory.
Both designs are clocked at $200$ MHz.
% We run \textit{Bert} on the baseline design and \textit{FNet} on our butterfly accelerator.
We evaluate both base ($12$ layers) and large ($24$ layers) versions of each model using four different input sequences ($128$, $256$, $512$ and $1024$).

A speedup breakdown is shown in~\figref{fig:breakdown_baseline}.
To demonstrate the improvement brought by our algorithm,
we first evaluate both \textit{BERT-Base} and \textit{FABNet} on the baseline design.
As the FFT is not supported in the baseline design,
we implement the Fourier layers as linear layers by multiplying the input sequences with DFT matrices.
Since the operation reduction brought by the algorithm is not fully utilized by the baseline design,
\textit{FABNet} results in a $1.6 \sim 2.3\times$ speedup compared to \textit{BERT}.
To further evaluate the improvement brought by hardware optimizations,
we evaluate \textit{FABNet} on our butterfly accelerator,
showing $19.5 \sim 53.3\times$ speedup when compared to the baseline design.
By combining both algorithm and hardware optimizations,
the overall speedup of our approach is $30.8 \sim 87.3\times$ over the baseline design.

\subsection{Comparison with GPU and CPU}\label{subsec:eval_cpugpu}

We compare our butterfly accelerator against GPU and CPU in both edge and server scenarios.
In the \textbf{edge scenario},
our butterfly accelerator is implemented on a Xilinx Zynq 7045 FPGA.
DDR4 is used as external memory and $512$ multipliers are used for computation.
Nvidia Jetson Nano GPU and Raspberry Pi4 are used as the GPU and CPU platforms, respectively.
In the \textbf{server scenario},
the butterfly accelerator is implemented on a Xilinx VCU128 FPGA.
HBM is used as external memory and the design consumes $1920$ multipliers.
We use Nvidia V100 and TITAN Xp GPUs for comparison, with highly-optimized CUDA implementations~\cite{dao2020kaleidoscope}.
FPGA designs are clocked at $200$ MHz.

We evaluate both \textit{FABNet-Base} and \textit{FABNet-Large} using $128$, $256$, $512$ and $1024$ input sequences.
\figref{fig:compare_cpu_gpu} shows the results in term of speedup and energy efficiency.
We represent energy efficiency using Giga operations per second per Watt (GOPS/Watt).
In the edge scenario,
our design on Zynq 7045 FPGA achieves $3.5 \sim 8\times$ speedup over Jetson Nano GPU and $36.6 \sim 342.3\times$ speedup over Raspberry Pi4\footnote{On \textit{FABNet-Large} with long input sequences greater than $768$,
Raspberry Pi 4 suffers from out-of-memory (OOM) issues.}.
At the same time,
our design yields $7.0 \sim 15.1\times$ and $162.8 \sim 273.8\times$ higher energy efficiency than Jetson Nano and Raspberry Pi4, respectively.
In the server scenario,
our design on VCU128 is up to $8.0$ and $9.0\times$ faster and up to $74.0$ and $79.4\times$ more energy-efficient than the V100 and TITAN Xp GPU, respectively.
In summary, the end-to-end speedup and energy efficiency gains on both edge and server scenarios under different input sequences highlight the scalability of our butterfly accelerator.

\begin{table}[t]
\centering
\caption{Hardware specification of CPU, GPU and FPGA.}
\vspace{2.0 mm}
\label{tb:hw_spec}
\setlength\tabcolsep{1pt} 
\scalebox{0.75}{
% \begin{tabular}{L{2cm}ccC{1.5cm}C{2.5cm}}
\begin{tabular}{C{1.0cm}|C{3.0cm}|C{1.3cm}|C{1.7cm}|C{1.9cm}|C{1.9cm}}
\toprule
& {\bf Platform}& {\bf \# cores}& {\bf Compiler} & {\bf Frequency} &  {\bf Technology}\\
\midrule
{\textbf{CPU}} & {Raspberry Pi 4} & 4 &  & {\SI{1.5}{\giga\hertz}} & {\SI{28}{\nano\meter}} \\
\cmidrule{1-3} \cmidrule{5-6}
 & {Nvidia V100} & 5,120 & {PyTorch} & {\SI{1.38}{\giga\hertz}} & {\SI{12}{\nano\meter}} \\
\cmidrule{2-3} \cmidrule{5-6}
 \textbf{GPU} & {Nvidia TITAN Xp} & 3,840 & {1.10.2} & {\SI{1.6}{\giga\hertz}} & {\SI{16}{\nano\meter}} \\
\cmidrule{2-3} \cmidrule{5-6}
 & {Nvidia Jetson Nano} & 128 & & {\SI{921.6}{\mega\hertz}} & {\SI{20}{\nano\meter}} \\
 \midrule
\multirow{2}{*}{\textbf{FPGA}} & {Xilinx VCU128} & - & Vivado & {\SI{200}{\mega\hertz}} & {\SI{16}{\nano\meter}} \\
\cmidrule{2-3} \cmidrule{5-6}
 & {Xlinx Zynq 7045} & - & 2019.2 & {\SI{200}{\mega\hertz}} & {\SI{28}{\nano\meter}} \\
\bottomrule
\end{tabular}}
% \vspace{-1.5mm}
\end{table}

\subsection{Comparison with SOTA Accelerators}\label{subsec:eval_sota}

\begin{table*}[htb]
\centering
\caption{Comparison with existing \textit{Transformer} accelerators in terms of latency, power and energy efficiency.}
\vspace{0.3cm}
\label{tb:sota_acc_perf}
\setlength\tabcolsep{1pt}
\scalebox{0.76}{
\begin{tabular}{C{3.3cm}|C{2.3cm}|C{2.3cm}|C{2.3cm}|C{2.3cm}|C{2.3cm}|C{2.5cm}|C{2.5cm}|C{2.5cm}}
\toprule
\multirow{2}{*}{ \textbf{Accelerators} } & $A^{3}$~\cite{ham2020} & SpAtten~\cite{wang2020spatten} & Sanger~\cite{sanger201micro} & Energon~\cite{zhou2021energon} & ELSA~\cite{elsa2021isca} & DOTA~\cite{dota2022asplos}&  FTRANS~\cite{li2020ftrans} & \multirow{2}{*}{Our work} \\ 
& (HPCA'20) & (HPCA'21)  & (MICRO'21) & (TCAD'21) & (ISCA'21) & (ASPLOS'22) & (ISLPED'20)  & \\ \midrule
\textbf{Technology} & ASIC (40nm) & ASIC (40nm) & ASIC (55nm) & ASIC (45nm) & ASIC (40nm) & ASIC (22nm) & FPGA (16nm) &  FPGA (16nm) \\ \midrule 
\textbf{Frequency} & \multicolumn{6}{c|}{1 GHz} & {170 MHz}  & 200 MHz\\ \midrule
\textbf{\# of Multipliers} & \multicolumn{6}{c|}{128} & {6531}  & 640 \\ \midrule
\textbf{Latency (ms)} & 56.0 & 48.8  & 45.2 & 44.2 & 34.7 & 34.1 & 61.6  & \textbf{2.4}\\ \midrule
\textbf{Throughput (Pred./s)} & 17.86 & 20.49 & 22.12 & 22.62 & 28.82  & 29.32 & 16.23 & \textbf{416.66}\\ \midrule
\textbf{Power (W)} & 1.217 & 1.060 & 0.801 & 2.633 & 0.976  & 0.858 & 25.130 &  11.355\\ \midrule
\textbf{Energy Eff. (Pred./J)} & 14.67 & 19.33 & 27.62 & 8.59 & 29.52  & 34.18 & 0.65 & \textbf{36.69}\\
\bottomrule
\end{tabular}
}
\end{table*}

\begin{figure*}[htb]
\centering
\includegraphics[width=0.99\textwidth]{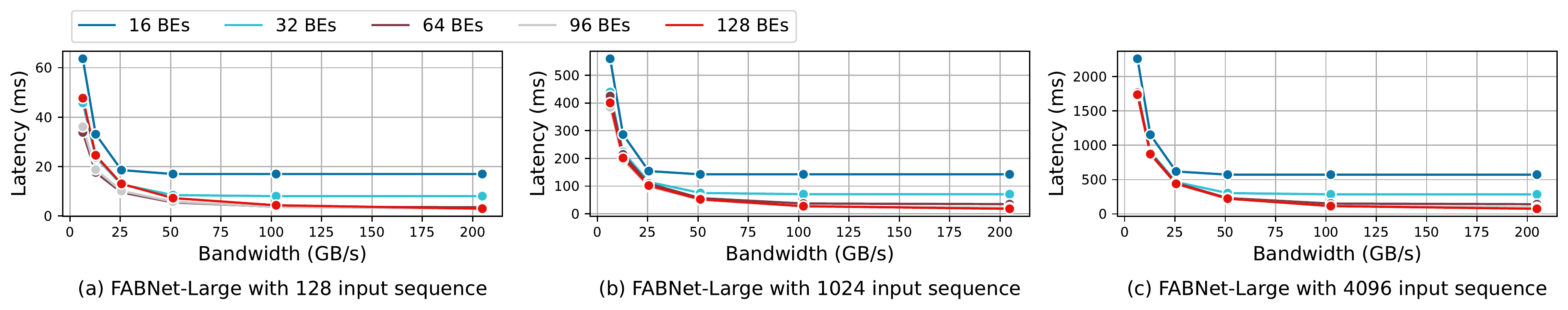}
% \vspace{-3mm}
\caption{{Latency for different input sequence lengths (a-c) when varying the available off-chip memory bandwidth.}}
\label{fig:bandwidth}
\end{figure*}

\tabref{tb:sota_acc_perf} compares our butterfly accelerator with existing state-of-the-art (SOTA) accelerators in terms of speed and power consumption.
Instead of comparing the effective throughput~\cite{wang2020spatten, sanger201micro},
we use the end-to-end latency to represent the actual execution speed of the hardware.
The energy efficiency is represented by the number of predictions per Joule (Pred./J).
Following the experimental setting of~\cite{dota2022asplos}, we compare all other SOTA accelerators on LRA-\textit{Image} dataset with one-layer vanilla \textit{Transformer}.
Among these accelerators,
only SpAtten~\cite{wang2020spatten} and 
DOTA~\cite{dota2022asplos} report the end-to-end performance.
For the rest of the accelerators that only support attention,
we estimate their performance by reusing their available multipliers to accelerate FFN.
Furthermore,
in both~\cite{wang2020spatten} and~\cite{sanger201micro},
the authors compare different ASIC and FPGA designs based on the assumption that all the ASIC designs are clocked at $1$ GHz with $128$ multipliers.
For a fair comparison,
we follow the same assumption in our experiments.
For designs with more than $128$ multipliers,
we follow the scaling approach of~\cite{wang2020spatten, sanger201micro} to linearly scale down its throughput to get their end-to-end performance.
For instance,
DOTA~\cite{dota2022asplos}\footnote{We assume their design is compute-bound.} achieves $11.4\times$ speedup over Nvidia V100 using $12,000$ multipliers with $12$ TOPS throughput. 
We scale down its throughput by $12,000/128=93.75$, which leads to $0.123\times$ speedup over V100.
To obtain the power consumption, we use the same linear scaling approach.
For instance,
Sanger~\cite{sanger201micro} reports the power consumption of a design with $1024$ multipliers.
We divide the power consumption of their systolic array ($2243$ mW) by $1024/128=8$, which leads to $280.375$ mW.
Together with the power of other modules such as pre-processing and memory, their total power consumption is $0.801$ W.
To match the computational capacity of ASIC designs,
we use $640$ DSPs in the VCU128 FPGA.
As our FPGA-based design is clocked at $200$ MHz,
this ensures that we have the same \mbox{$640 \times 200$M $= 1$28 GOPS} theoretical peak performance as ASIC designs (\mbox{$128 \times 1$G $= 1$28 GOPS}).
While this is a simple approximation,
it allows us to compare different hardware architectures regardless of their underlying target platforms.

As shown in~\tabref{tb:sota_acc_perf},
our butterfly accelerator achieves $25.6\times$ speedup over the FPGA-based FTRANS~\cite{li2020ftrans} while using nearly $10\times$ fewer DSPs.
At the same time,
we achieve $62.3\times$ higher energy efficiency than FTRANS.
Compared with ASIC designs,
our accelerator achieves $14.2 \sim 23.2\times$ speedup under the same computational capacity.
Although our FPGA-based butterfly design consumes more power than ASIC designs,
it yields $1.1 \sim 4.3\times$ higher energy efficiency than the other SOTA ASIC accelerators.
We expect further speedup and energy efficiency improvements when our design is implemented as an ASIC.

We attribute the performance gain of our approach over ASIC designs to two main factors:
\textit{1)}~the use of FFT and butterfly factorization which significantly reduces the computational complexity at the algorithmic level; \textit{2)}~the adaptable butterfly design that adopts a single unified hardware engine to accelerate both FFT and butterfly linear transformation, which significantly improves the hardware efficiency; and \textit{3)}~the co-design process which jointly optimizes both algorithm and hardware parameters.

\subsection{Off-Chip Memory Bandwidth Analysis}\label{subsec:eval_bandwidth}

In order to investigate the sensitivity of our design to off-chip memory bandwidth,
we vary the bandwidth from $6$, $12$, $25$, $50$, $100$ and $200$ GB/s, and evaluate its latency based on our performance model.
For these experiments, we use five different designs with $16$, $32$, $64$ and $128$ \textit{BE}s executing \textit{FABNet-Large} with $24$ layers.
To understand the bandwidth requirements under both short and long input lengths, we evaluate each design using three input sequences ($128$, $1024$ and $4096$).
The results are shown in~\figref{fig:bandwidth}.
For a small-scale design of $16$ \textit{BE}s, a bandwidth of $50$ GB/s is enough for the design to reach its peak performance under different input sequences.
For the largest design of $128$ \textit{BE}s,
the achieved performance saturates once the bandwidth reaches $100$ GB/s.

\subsection{Power and Resource Analysis}

\tabref{tb:power_breakdown} shows the  power consumption breakdown\footnote{\small Power of I/O is not included as it occupies less than $1$\% of the total power.} based on the report generated from the Vivado XPE tool.
We implement two designs with $120$ \textit{BEs} (\textit{BE-120}) and $40$ \textit{BEs} (\textit{BE-40}) on a VCU128 FPGA, which have been used in~\secref{subsec:eval_cpugpu} and~\secref{subsec:eval_sota}, respectively.
In both designs,
the dynamic power accounts for more than $70$\% of the total power consumption.
The memory resources,
including both BRAM and HBM,
consume more than $25$\% of the dynamic power.
Furthermore, when the number of \textit{BEs} scales from $40$ to $120$, the power of clocking, logic \& signal and DSPs increases from $2.688$~W, $2.381$~W and $0.338$~W to $6.882$~W, $7.732$~W and $1.437$~W, respectively.

\tabref{tb:resource_breakdown} presents the resource consumption of both \textit{BE-40} and \textit{BE-120} designs on the same VCU120 FPGA.
Due to the use of FFT and butterfly matrices,
our \textit{FABNet} becomes less memory-intensive than the vanilla attention-based NNs.
Since the theoretical memory bandwidth of a single HBM ($450$~GB/s) can already satisfy the requirement of our accelerator~(\secref{subsec:eval_bandwidth}),
we use one HBM in both designs to reduce the resource and power consumption.
When the number of \textit{BE}s decreases from $120$ to $40$,
the BRAM usage is reduced from $978$ to $338$.
This reduction can also be observed on the LUT and register resources.

\begin{table}[htb]
\centering
% \vspace{-2.mm}
\caption{Power breakdown of our designs on VCU128.}
\vspace{2.0 mm}
\label{tb:power_breakdown}
\setlength\tabcolsep{1pt} 
\scalebox{0.78}{
% \begin{tabular}{L{2cm}ccC{1.5cm}C{2.5cm}}
\begin{tabular}{C{1.2cm}|C{1.0cm}|C{1.5cm}|C{1.5cm}|C{1.1cm}|C{2.8cm}|C{1.2cm}}
\toprule
\multicolumn{2}{c|}{}&\multicolumn{4}{c|}{\textbf{Dynamic} (W)} & \multirow{3}{*}{\textbf{Static}} \\ \cmidrule{3-6}
\multicolumn{2}{c|}{\textbf{Design}}& \multirow{2}{*}{Clocking} &  {Logic\&} &  \multirow{2}{*}{DSP} & {Memory} & \\ 
\multicolumn{2}{c|}{}&  & {Signal} &   & (BRAM \& HBM) & (W) \\ \midrule
\multirow{3}{*}{\textit{\textbf{BE-40}}}& used & 2.668 & 2.381 & 0.338 & 5.325 & 3.368 \\ \cmidrule{2-7}
& pct. & 18.8\% & 16.7\% & 2.3\%  & 37.5\% & 23.7\%\\ \midrule
\multirow{3}{*}{\textit{\textbf{BE-120}}}& used & 6.882 & 7.732 & 1.437 & 6.142 & 3.665\\ \cmidrule{2-7}
& pct. & 26.4\%  & 29.7\% & 5.5\% & 23.6\%  & 14.1\% \\
\bottomrule
\end{tabular}}
% \vspace{-4.5mm}
\end{table}

\begin{table}[htb]
\centering
\caption{Resource usage of our designs on VCU128.}
\vspace{2.0 mm}
\label{tb:resource_breakdown}
\setlength\tabcolsep{1pt} 
\scalebox{0.8}{
\begin{tabular}{C{1.2cm}|C{1.0cm}|C{1.7cm}|C{1.7cm}|C{1.5cm}|C{1.6cm}|C{1.2cm}}
\toprule
\multicolumn{2}{c|}{}& \textbf{LUTs} & \textbf{Registers} & \textbf{DSP48s} & \textbf{BRAMs} & \textbf{HBMs} \\ \midrule
\multicolumn{2}{c|}{\textbf{Available}} & 1,303,680 & 2,607,360 & 9,024 & 2,016 & 2 \\ \midrule
\multirow{3}{*}{\textbf{BE-40}}& used & 358,609  & 536,810 & 640 & 338 & 1  \\ \cmidrule{2-7}
& pct. & 27.5\% & 20.6\% & 7.1\%  & 16.8\% & 50.0\% \\ \midrule
\multirow{3}{*}{\textbf{BE-120}}& used & 1,034,610 & 1,648,695 & 2,880 & 978 & 1  \\ \cmidrule{2-7}
& pct. & 79.3\% & 63.2\% & 31.9\%  & 48.5\% & 50.0\% \\
\bottomrule
\end{tabular}}
% \vspace{-1.5mm}
\end{table}

%% file: 7_related_work.tex
\section{Related Work}\label{sec:background}

\noindent
\textbf{Efficient Approaches for Attention.}
As the algorithmic complexity of the self-attention mechanism scales quadratically with respect to the input sequence length,
many sparse variants have been introduced to approximate the attention-based NNs~\cite{tay2020efficient}.
The sparsity patterns in these approaches are determined either dynamically~\cite{wang2020linformer, zaheer2020big, choromanski2020rethinking, tay2020sparse} or statically~\cite{beltagy2020longformer,lee2021fnet,chen2021pixelated}. 
Although these methods achieve high compression rate on the number of operations and parameters,
the hardware cost and efficiency of their mappings on real hardware designs are not considered in these works.

\noindent
\textbf{Domain-Specific Accelerators for Attention-based NNs.}
To better utilize existing efficient attention-based algorithmic approaches on hardware, various domain-specific hardware designs have been introduced.
Ham \textit{et al.}~\cite{ham2020} propose a hardware architecture called $A^3$, which dynamically prunes entries based on their softmax importance.
% With the customized hardware design and algorithmic approximation,
% $A^3$ can achieve nearly $100 \times$ speedup compared with the implementation on TITAN Xp GPU.
% However, their design only focuses on optimizing the attention mechanism without accelerating FFN. 
% To exploit the redundancy in the attention mechanism,
By leveraging the sparsity in both head and token levels,
Wang~\textit{et al.}~\cite{wang2020spatten} propose SpAtten that dynamically prunes entire rows and columns from the attention matrix.
EdgeBERT~\cite{edgebert2021micro} explores the layer sparsity via an entropy-based early-exit approach, which significantly reduces the computation and memory footprint.
To detect the weak relationship in attention, DOTA~\cite{dota2022asplos} uses low-rank linear transformations to detect and omit the weak connections.
ELSA~\cite{elsa2021isca} approximates the attention mechanism using a sign random projection approach.
To further exploit the sparsity in attention-based NNs,
Energon~\cite{zhou2021energon} adopts a low-precision NN to predict the sparsity in the attention matrix. 
However, the generated sparsity patterns from these approaches are always unstructured, which may lead to hardware inefficiency.
Sanger~\cite{sanger201micro} propose pack-and-split modules to distribute the non-zero computation to each computation engine, achieving a load-balanced execution.
Although these accelerators achieve notable speedup over GPUs and CPUs when executing the attention mechanism,
their end-to-end hardware performance is limited as
the approximation and acceleration of the FFN part are not considered in their design.

\noindent
\textbf{Comparison to Previous Work.}
As spatial-domain convolution corresponds to frequency-domain multiplication,
various FFT-based hardware accelerators have been introduced to accelerate CNNs~\cite{zhang2017frequency, abtahi2018fft}, LSTMs~\cite{wang2018c, li2019rnn} and attention-based NNs~\cite{li2020ftrans}.
However, these designs only use FFT as a domain transfer approach, while the main computation is still performed in another processing engine.
In contrast,
this paper adopts the butterfly sparsity, a generalized pattern of FFT, for the main computation of the network.
Based on the proposed method, all the computations are performed in a single unified butterfly accelerator, resulting in a higher hardware efficiency over previous designs.

Although butterfly sparsity has been explored in recent literature,
most approaches only focus on algorithm-level optimizations.
Dao~\textit{et al.}~\cite{dao2020kaleidoscope} demonstrate the potential of the butterfly matrix in approximating linear transformations, but its efficiency on attention matrices is not explored in their work.
On the other hand, Pixelated Butterfly~\cite{chen2021pixelated} and Sparse Transformer~\cite{child2019generating} focus on adopting the butterfly pattern for attention matrices, neglecting the linear layers. 
Moreover, their designs require the use of multiple sparsity patterns to compensate for the accuracy loss, which significantly complicates the hardware design.
Lee-Thorp~\textit{et al.}~\cite{lee2021fnet} show the effectiveness of Fourier transforms in accelerating attention layers, but did not consider the linear layers, leading to scalability issues~(\secref{subsec:motivation}).
Different from all these efforts,
this paper exploits the use of butterfly sparsity for both attention and linear layers via an algorithm-hardware co-design approach.
A novel butterfly-based algorithm and an adaptable hardware accelerator are jointly designed to overcome existing limitations to push the performance limit.

%% file: 8_conclusion.tex
\section{Conclusion}
This paper proposes the end-to-end acceleration of attention-based NNs via algorithm and hardware co-design.
On the algorithmic level,
we propose \textit{FABNet}, a hardware-friendly attention-based NN.
Both attention and linear layers are compressed using a unified butterfly sparsity pattern allowing for scalable end-to-end acceleration.
On the hardware level,
an adaptable butterfly accelerator is proposed that can be configured at runtime to accelerate different layers based on a unified hardware engine to achieve high hardware efficiency.
Both algorithm and hardware design parameters are jointly optimized to push the performance limit.
Our experiments demonstrate that our co-design approach yields up to $23.2\times$ speedup over state-of-the-art accelerators. Furthermore, our design achieves up to $273.8\times$ higher energy efficiency compared to optimized GPU implementations.

\section*{Acknowledgement}
The support of the UK EPSRC grants (UK EPSRC grant number EP\slash V028251\slash 1, EP\slash L016796\slash 1, EP\slash S030069\slash 1 and EP\slash N031768\slash 1), AMD and Intel is gratefully acknowledged.
We also thank Alexander Mathiasen for insightful discussions on model compression.

%% file: 9_artifact_evaluation.tex
% LaTeX template for Artifact Evaluation V20201122
%
% Prepared by 
% * Grigori Fursin (cTuning foundation, France) 2014-2020
% * Bruce Childers (University of Pittsburgh, USA) 2014
%
% See examples of this Artifact Appendix in
%  * SC'17 paper: https://dl.acm.org/citation.cfm?id=3126948
%  * CGO'17 paper: https://www.cl.cam.ac.uk/~sa614/papers/Software-Prefetching-CGO2017.pdf
%  * ACM ReQuEST-ASPLOS'18 paper: https://dl.acm.org/citation.cfm?doid=3229762.3229763
%
% (C)opyright 2014-2020
%
% CC BY 4.0 license
%

%%%%%%%%%%%%%%%%%%%%%%%%%%%%%%%%%%%%%%%%%%%%%%%%%%%%
% When adding this appendix to your paper, 
% please remove above part
%%%%%%%%%%%%%%%%%%%%%%%%%%%%%%%%%%%%%%%%%%%%%%%%%%%%
\appendix
\section{Artifact Appendix}

%%%%%%%%%%%%%%%%%%%%%%%%%%%%%%%%%%%%%%%%%%%%%%%%%%%%%%%%%%%%%%%%%%%%%
\subsection{Abstract}
This Appendix summarizes the necessary information and instructions to evaluate our artifacts.
The functionality of our hardware accelerator can be evaluated by running Verilog HDL designs and System Verilog testbenches on Vivado design suite.
The accuracy results can be obtained by running our PyTorch programs and the associated Bash scripts.
The power and resource utilization can be obtained by running Synthesis and Implementation using our RTL code and constraint files.
The latency can be obtained by running our custom Python-based performance model.
We also provide all our training log files and Vivado design reports in the link: \url{https://drive.google.com/drive/folders/1jaR8gDX-zO1Hu83xFg_IJOwRgoBMPnjY?usp=sharing}.

\subsection{Artifact check-list (meta-information)}

\begin{itemize}
  \item {\bf Algorithm: } \textit{FABNet}, an efficient model that adopts a unified butterfly sparsity pattern to approximate both the attention mechanism and the FFNs.
  \item {\bf Program: } Python, PyTorch, Verilog HDL
  \item {\bf Model: } We evaluate three models for comparison, including \textit{Transformer}, \textit{FNet} and \textit{FABNet}.
  \item {\bf Data set: } Long-Range-Arena~(LRA) dataset, which is a well-known long sequence natural language processing (NLP) dataset. The zip file can be downloaded from the link: \url{https://storage.googleapis.com/long-range-arena/lra_release.gz}. The required disk space of the unzip file is around 33 GB.
  \item {\bf Run-time environment: } Ubuntu 20.04, CUDA SDK 11.3 or higher.
  \item {\bf Hardware: } Nvidia V100 GPU, Nvidia TITAN Xp GPU, Nvidia Jetson Nano GPU, Intel Xeon Gold 6154 CPU, Raspberry Pi 4.
  \item {\bf Metrics: } Accuracy, simulated latency, resource and power consumption.
  \item {\bf Experiments: } Bash scripts and detailed instructions are provided to run experiments.
  \item {\bf How much disk space required (approximately)?} 
  \item {\bf How much time is needed to prepare workflow (approximately)?} $1\sim2$ hours.
  \item {\bf How much time is needed to complete experiments (approximately)?} Accuracy results: hundreds of GPU hours to obtain. Power and resource consumption: around 70 hours. Functionality of Verilog design: around 5 hours.
  \item {\bf Publicly available?} Yes.
  \item {\bf Code licenses (if publicly available)? } Yes.
  \item {\bf Archived (provide DOI)?} We will update this later.
\end{itemize}

%%%%%%%%%%%%%%%%%%%%%%%%%%%%%%%%%%%%%%%%%%%%%%%%%%%%%%%%%%%%%%%%%%%%%
\subsection{Description}\label{sec:ae}
\subsubsection{How to access}
You can access our codebase from the link:
\url{https://zenodo.org/record/7010800#.YwQKCOzMJhF} or \url{https://github.com/os-hxfan/Butterfly_Acc.git}.

\subsubsection{Hardware dependencies} A GPU server is required to run the training of our models.
A CPU server is needed to run simulation, synthesis and place\&route. Different GPUs and CPUs, such as Nvidia Jetson Nano GPU and Raspberry Pi 4, are also required to evaluate the hardware performance of different models.

\subsubsection{Software dependencies}  Vivado Design Suite $2019.2$, PyTorch $1.10.2$, CUDA SDK $11.3$ or higher, Python $3.8$ or higher. Other dependencies are listed in \textbf{requirements.txt}.

\subsubsection{Data sets}
Five tasks in the LRA dataset including \textit{ListOPs} for hierarchical data classification, \textit{Text} for byte-level text classification, \textit{Retrieval} for byte-level document retrieval, \textit{Image} for image classification for sequences of pixels and \textit{Pathfinder} for  classification of long-range spatial dependency.

%%%%%%%%%%%%%%%%%%%%%%%%%%%%%%%%%%%%%%%%%%%%%%%%%%%%%%%%%%%%%%%%%%%%%
\subsection{Installation}

We provide a detailed installation guide in the {\path{README.md}} of the root directory.
%%%%%%%%%%%%%%%%%%%%%%%%%%%%%%%%%%%%%%%%%%%%%%%%%%%%%%%%%%%%%%%%%%%%%
\subsection{Experiment workflow}\label{subsec:exp_flow}
To evaluate the functionality of our hardware, perform the following steps:
\begin{itemize}
    \item Follow the instruction of experimental setup in the root directory to install software dependencies. Install Vivado 2019.2. 
    \item Generate the test data using our Python programs. 
    \item Create a Vivado project for our hardware design.
    \item Import all the Verilog source code and System Verilog testbenches.
    \item Include all the necessary IPs from Vivado IP library.
\end{itemize}
We provide Vivado Tcl scripts and step-by-step instructions in \path{./hardware/npu_design/verilog/README.md} to automate the whole process.

To reproduce the algorithmic and hardware performance, we provide all the scripts under the directory \path{./script_figs} to generate figures and tables. The detailed instructions are provided in \path{./script_figs/README.md}.

%%%%%%%%%%%%%%%%%%%%%%%%%%%%%%%%%%%%%%%%%%%%%%%%%%%%%%%%%%%%%%%%%%%%%
\subsection{Evaluation and expected results}
We provide scripts under \path{./script_figs} to generate all the figures and tables related to accuracy performance and hardware performance include power consumption, resource utilization and simulated latency performance.
As running all the experiments requires a few hundred GPU/CPU hours, to facilitate the artifact evaluation,
we refer to the following key results that can be obtained within a reasonable time:
\begin{itemize}
    \item Vivado simulation to run different layers, such as fast Fourier transform, butterfly matrix multiplication and layer normalization, on our Verilog hardware design. We provide System Verilog testbenches under \path{./hardware/npu_design/verilog/functionality/testbench/}, and a detailed workflow in the first paragraph of~\secref{subsec:exp_flow}.
    \item Power breakdown in~\tabref{tb:power_breakdown} and resource utilization in~\tabref{tb:resource_breakdown}. We provide detailed instructions and Vivado Tcl scripts under \path{./hardware/npu_design/verilog/} to run synthesis and place\&route on both VCU128 and Zynq 7045 FPGAs.  
\end{itemize}

Although it takes longer to run other experiments, all the results are reproducible using our provided scripts.
We provide all the GPU training log files and Vivado design reports in the link: \url{https://drive.google.com/drive/folders/1jaR8gDX-zO1Hu83xFg_IJOwRgoBMPnjY?usp=sharing}.
%%%%%%%%%%%%%%%%%%%%%%%%%%%%%%%%%%%%%%%%%%%%%%%%%%%%%%%%%%%%%%%%%%%%%
\subsection{Methodology}

Submission, reviewing and badging methodology:

\begin{itemize}
  \item \url{https://www.acm.org/publications/policies/artifact-review-badging}
  \item \url{http://cTuning.org/ae/submission-20201122.html}
  \item \url{http://cTuning.org/ae/reviewing-20201122.html}
\end{itemize}

%% file: main.bbl
\begin{thebibliography}{10}
\providecommand{\url}[1]{#1}
\csname url@samestyle\endcsname
\providecommand{\newblock}{\relax}
\providecommand{\bibinfo}[2]{#2}
\providecommand{\BIBentrySTDinterwordspacing}{\spaceskip=0pt\relax}
\providecommand{\BIBentryALTinterwordstretchfactor}{4}
\providecommand{\BIBentryALTinterwordspacing}{\spaceskip=\fontdimen2\font plus
\BIBentryALTinterwordstretchfactor\fontdimen3\font minus
  \fontdimen4\font\relax}
\providecommand{\BIBforeignlanguage}[2]{{%
\expandafter\ifx\csname l@#1\endcsname\relax
\typeout{** WARNING: IEEEtran.bst: No hyphenation pattern has been}%
\typeout{** loaded for the language `#1'. Using the pattern for}%
\typeout{** the default language instead.}%
\else
\language=\csname l@#1\endcsname
\fi
#2}}
\providecommand{\BIBdecl}{\relax}
\BIBdecl

\bibitem{brauwers2021general}
G.~Brauwers and F.~Frasincar, ``{A General Survey on Attention Mechanisms in
  Deep Learning},'' \emph{IEEE Transactions on Knowledge and Data Engineering},
  2021.

\bibitem{vaswani2017attention}
A.~Vaswani, N.~Shazeer, N.~Parmar, J.~Uszkoreit, L.~Jones, A.~N. Gomez,
  L.~Kaiser, and I.~Polosukhin, ``{Attention is All You Need},'' \emph{Advances
  in Neural Information Processing Systems (NeurIPS)}, 2017.

\bibitem{devlin2018bert}
J.~Devlin, M.-W. Chang, K.~Lee, and K.~Toutanova, ``{BERT: Pre-training of Deep
  Bidirectional Transformers for Language Understanding},'' in \emph{ACL},
  2019.

\bibitem{radford2019language}
A.~Radford, J.~Wu, R.~Child, D.~Luan, D.~Amodei, and I.~Sutskever, ``{Language
  Models are Unsupervised Multitask Learners},'' \emph{OpenAI blog}, vol.~1,
  no.~8, p.~9, 2019.

\bibitem{dosovitskiy2020image}
A.~Dosovitskiy, L.~Beyer, A.~Kolesnikov, D.~Weissenborn, X.~Zhai,
  T.~Unterthiner, M.~Dehghani, M.~Minderer, G.~Heigold, S.~Gelly \emph{et~al.},
  ``{An image is worth 16x16 words: Transformers for image recognition at
  scale},'' in \emph{International Conference on Representation Learning
  (ICLR)}, 2021.

\bibitem{wang2020spatten}
H.~Wang, Z.~Zhang, and S.~Han, ``{SpAtten: Efficient Sparse Attention
  Architecture with Cascade Token and Head Pruning},'' \emph{IEEE International
  Symposium on High Performance Computer Architecture (HPCA)}, 2021.

\bibitem{choromanski2020rethinking}
K.~Choromanski, V.~Likhosherstov, D.~Dohan, X.~Song, A.~Gane, T.~Sarlos,
  P.~Hawkins, J.~Davis, A.~Mohiuddin, L.~Kaiser \emph{et~al.}, ``{Rethinking
  Attention with Performers},'' in \emph{International Conference on Learning
  Representations (ICLR)}, 2021.

\bibitem{wang2020linformer}
S.~Wang, B.~Z. Li, M.~Khabsa, H.~Fang, and H.~Ma, ``{Linformer: Self-Attention
  with Linear Complexity},'' \emph{arXiv preprint arXiv:2006.04768}, 2020.

\bibitem{kitaev2020reformer}
N.~Kitaev, {\L}.~Kaiser, and A.~Levskaya, ``{Reformer: The Efficient
  Transformer},'' in \emph{International Conference on Learning Representations
  (ICLR)}, 2020.

\bibitem{tay2020sparse}
Y.~Tay, D.~Bahri, L.~Yang, D.~Metzler, and D.-C. Juan, ``{Sparse Sinkhorn
  Attention},'' in \emph{International Conference on Machine Learning (ICML)},
  2020.

\bibitem{beltagy2020longformer}
I.~Beltagy, M.~E. Peters, and A.~Cohan, ``{Longformer: The Long-Document
  Transformer},'' \emph{arXiv preprint arXiv:2004.05150}, 2020.

\bibitem{zaheer2020big}
M.~Zaheer, G.~Guruganesh, K.~A. Dubey, J.~Ainslie, C.~Alberti, S.~Ontanon,
  P.~Pham, A.~Ravula, Q.~Wang, L.~Yang \emph{et~al.}, ``{Big Bird: Transformers
  for Longer Sequences},'' in \emph{Advances in Neural Information Processing
  Systems (NeurIPS)}, 2020.

\bibitem{child2019generating}
R.~Child, S.~Gray, A.~Radford, and I.~Sutskever, ``{Generating Long Sequences
  with Sparse Transformers},'' \emph{arXiv preprint arXiv:1904.10509}, 2019.

\bibitem{ham2020}
T.~J. Ham, S.~J. Jung, S.~Kim, Y.~H. Oh, Y.~Park, Y.~Song, J.-H. Park, S.~Lee,
  K.~Park, J.~W. Lee \emph{et~al.}, ``{A$^{\text{3}}$: Accelerating Attention
  Mechanisms in Neural Networks with Approximation},'' in \emph{IEEE
  International Symposium on High Performance Computer Architecture (HPCA)},
  2020.

\bibitem{sanger201micro}
L.~Lu, Y.~Jin, H.~Bi, Z.~Luo, P.~Li, T.~Wang, and Y.~Liang, ``{Sanger: A
  Co-Design Framework for Enabling Sparse Attention Using Reconfigurable
  Architecture},'' in \emph{54th Annual IEEE/ACM International Symposium on
  Microarchitecture (MICRO)}, 2021.

\bibitem{zhou2021energon}
Z.~Zhou, J.~Liu, Z.~Gu, and G.~Sun, ``{Energon: Towards Efficient Acceleration
  of Transformers Using Dynamic Sparse Attention},'' \emph{IEEE Transactions on
  Computer-Aided Design of Integrated Circuits and Systems (TCAD)}, 2021.

\bibitem{elsa2021isca}
T.~J. Ham, Y.~Lee, S.~H. Seo, S.~Kim, H.~Choi, S.~J. Jung, and J.~W. Lee,
  ``{ELSA: Hardware-Software Co-Design for Efficient, Lightweight
  Self-Attention Mechanism in Neural Networks},'' in \emph{ACM/IEEE 48th Annual
  International Symposium on Computer Architecture (ISCA)}, 2021.

\bibitem{dota2022asplos}
Z.~Qu, L.~Liu, F.~Tu, Z.~Chen, Y.~Ding, and Y.~Xie, ``{DOTA: Detect and Omit
  Weak Attentions for Scalable Transformer Acceleration},'' in \emph{27th ACM
  International Conference on Architectural Support for Programming Languages
  and Operating Systems (ASPLOS)}, 2022.

\bibitem{li2020ftrans}
B.~Li, S.~Pandey, H.~Fang, Y.~Lyv, J.~Li, J.~Chen, M.~Xie, L.~Wan, H.~Liu, and
  C.~Ding, ``{FTRANS: Energy-Efficient Acceleration of Transformers using
  FPGA},'' in \emph{ACM/IEEE International Symposium on Low Power Electronics
  and Design (ISLPED)}, 2020, pp. 175--180.

\bibitem{edgebert2021micro}
T.~Tambe, C.~Hooper, L.~Pentecost, T.~Jia, E.-Y. Yang, M.~Donato, V.~Sanh,
  P.~Whatmough, A.~M. Rush, D.~Brooks \emph{et~al.}, ``{EdgeBERT:
  Sentence-Level Energy Optimizations for Latency-Aware Multi-Task NLP
  Inference},'' in \emph{54th Annual IEEE/ACM International Symposium on
  Microarchitecture (MICRO)}, 2021.

\bibitem{jang2021sparsity}
J.-W. Jang, S.~Lee, D.~Kim, H.~Park, A.~S. Ardestani, Y.~Choi, C.~Kim, Y.~Kim,
  H.~Yu, H.~Abdel-Aziz \emph{et~al.}, ``{Sparsity-Aware and Re-configurable NPU
  Architecture for Samsung Flagship Mobile SoC},'' in \emph{ACM/IEEE 48th
  Annual International Symposium on Computer Architecture (ISCA)}, 2021.

\bibitem{he2016deep}
K.~He, X.~Zhang, S.~Ren, and J.~Sun, ``{Deep Residual Learning for Image
  Recognition},'' in \emph{IEEE Conference on Computer Vision and Pattern
  Recognition (CVPR)}, 2016, pp. 770--778.

\bibitem{ba2016layer}
J.~L. Ba, J.~R. Kiros, and G.~E. Hinton, ``{Layer Normalization},'' \emph{arXiv
  preprint arXiv:1607.06450}, 2016.

\bibitem{lample2019cross}
G.~Lample and A.~Conneau, ``Cross-lingual language model pretraining,''
  \emph{Advances in Neural Information Processing Systems (NeurIPS)}, 2019.

\bibitem{wang2018glue}
A.~Wang, A.~Singh, J.~Michael, F.~Hill, O.~Levy, and S.~R. Bowman, ``{GLUE: A
  Multi-Task Benchmark and Analysis Platform for Natural Language
  Understanding},'' \emph{Proceedings of the 2018 {EMNLP} Workshop
  {B}lackbox{NLP}: Analyzing and Interpreting Neural Networks for {NLP}}, 2018.

\bibitem{ma2019tensorized}
X.~Ma, P.~Zhang, S.~Zhang, N.~Duan, Y.~Hou, M.~Zhou, and D.~Song, ``{A
  Tensorized Transformer for Language Modeling},'' in \emph{Advances in Neural
  Information Processing Systems (NeurIPS)}, 2019.

\bibitem{chen2021scatterbrain}
B.~Chen, T.~Dao, E.~Winsor, Z.~Song, A.~Rudra, and C.~R\'{e}, ``{Scatterbrain:
  Unifying Sparse and Low-rank Attention},'' in \emph{Advances in Neural
  Information Processing Systems (NeurIPS)}, 2021.

\bibitem{parker1995random}
D.~S. Parker, ``{Random Butterfly Transformations with Applications in
  Computational Linear Algebra},'' 1995.

\bibitem{dao2019learning}
T.~Dao, A.~Gu, M.~Eichhorn, A.~Rudra, and C.~R{\'e}, ``{Learning Fast
  Algorithms for Linear Transforms Using Butterfly Factorizations},'' in
  \emph{International Conference on Machine Learning (ICML)}, 2019.

\bibitem{de2018two}
C.~De~Sa, A.~Cu, R.~Puttagunta, C.~R{\'e}, and A.~Rudra, ``A two-pronged
  progress in structured dense matrix vector multiplication,'' in
  \emph{Proceedings of the Twenty-Ninth Annual ACM-SIAM Symposium on Discrete
  Algorithms}.\hskip 1em plus 0.5em minus 0.4em\relax SIAM, 2018, pp.
  1060--1079.

\bibitem{chen2021pixelated}
B.~Chen, T.~Dao, K.~Liang, J.~Yang, Z.~Song, A.~Rudra, and C.~Re, ``{Pixelated
  Butterfly: Simple and Efficient Sparse training for Neural Network Models},''
  in \emph{International Conference on Learning Representations (ICLR)}, 2022.

\bibitem{dao2020kaleidoscope}
T.~Dao, N.~S. Sohoni, A.~Gu, M.~Eichhorn, A.~Blonder, M.~Leszczynski, A.~Rudra,
  and C.~R{\'e}, ``{Kaleidoscope: An Efficient, Learnable Representation for
  All Structured Linear Maps},'' in \emph{International Conference on Learning
  Representations (ICLR)}, 2020.

\bibitem{tolstikhin2021mlp}
I.~O. Tolstikhin, N.~Houlsby, A.~Kolesnikov, L.~Beyer, X.~Zhai, T.~Unterthiner,
  J.~Yung, A.~Steiner, D.~Keysers, J.~Uszkoreit \emph{et~al.}, ``{MLP-Mixer: An
  all-MLP Architecture for Vision},'' \emph{Advances in Neural Information
  Processing Systems (NeurIPS)}, 2021.

\bibitem{lee2021fnet}
J.~Lee-Thorp, J.~Ainslie, I.~Eckstein, and S.~Ontanon, ``{FNet: Mixing Tokens
  with Fourier Transforms},'' \emph{arXiv preprint arXiv:2105.03824}, 2021.

\bibitem{cooley1965algorithm}
J.~W. Cooley and J.~W. Tukey, ``An algorithm for the machine calculation of
  complex fourier series,'' \emph{Mathematics of computation}, vol.~19, no.~90,
  pp. 297--301, 1965.

\bibitem{amdahl1967validity}
G.~M. Amdahl, ``Validity of the single processor approach to achieving large
  scale computing capabilities,'' in \emph{Proceedings of the April 18-20,
  1967, Spring Joint Computer Conference}, 1967, pp. 483--485.

\bibitem{dao2022monarch}
T.~Dao, B.~Chen, N.~S. Sohoni, A.~Desai, M.~Poli, J.~Grogan, A.~Liu, A.~Rao,
  A.~Rudra, and C.~R{\'e}, ``{Monarch: Expressive Structured Matrices for
  Efficient and Accurate Training},'' in \emph{International Conference on
  Machine Learning (ICML)}, 2022, pp. 4690--4721.

\bibitem{geng2020awb}
T.~Geng, A.~Li, R.~Shi, C.~Wu, T.~Wang, Y.~Li, P.~Haghi, A.~Tumeo, S.~Che,
  S.~Reinhardt \emph{et~al.}, ``{AWB-GCN: A Graph Convolutional Network
  Accelerator with Runtime Workload Rebalancing},'' in \emph{53rd Annual
  IEEE/ACM International Symposium on Microarchitecture (MICRO)}, 2020.

\bibitem{tay2020long}
Y.~Tay, M.~Dehghani, S.~Abnar, Y.~Shen, D.~Bahri, P.~Pham, J.~Rao, L.~Yang,
  S.~Ruder, and D.~Metzler, ``{Long Range Arena: A Benchmark for Efficient
  Transformers},'' in \emph{International Conference on Learning
  Representations (ICLR)}, 2020.

\bibitem{pytorch}
A.~Paszke \emph{et~al.}, ``{PyTorch: An Imperative Style, High-Performance Deep
  Learning Library},'' in \emph{Advances in Neural Information Processing
  Systems (NeurIPS)}, 2019.

\bibitem{wolf2019huggingface}
T.~Wolf \emph{et~al.}, ``{Huggingface's Transformers: State-of-the-art Natural
  Language Processing},'' \emph{arXiv preprint arXiv:1910.03771}, 2019.

\bibitem{xiong2021nystromformer}
Y.~Xiong, Z.~Zeng, R.~Chakraborty, M.~Tan, G.~Fung, Y.~Li, and V.~Singh,
  ``{Nystr{\"o}mformer: A Nyst{\"o}m-based Algorithm for Approximating
  Self-Attention},'' in \emph{Proceedings of the AAAI Conference on Artificial
  Intelligence (AAAI)}, vol.~35, no.~16.\hskip 1em plus 0.5em minus 0.4em\relax
  NIH Public Access, 2021, p. 14138.

\bibitem{song2019hypar}
L.~Song, J.~Mao, Y.~Zhuo, X.~Qian, H.~Li, and Y.~Chen, ``{HyPar: Towards Hybrid
  Parallelism for Deep Learning Accelerator Array},'' in \emph{2019 IEEE
  International Symposium on High Performance Computer Architecture (HPCA)},
  2019.

\bibitem{alwani2016fused}
M.~Alwani, H.~Chen, M.~Ferdman, and P.~Milder, ``{Fused-Layer CNN
  Accelerators},'' in \emph{49th Annual IEEE/ACM International Symposium on
  Microarchitecture (MICRO)}, 2016.

\bibitem{tay2020efficient}
Y.~Tay, M.~Dehghani, D.~Bahri, and D.~Metzler, ``{Efficient Transformers: A
  Survey},'' \emph{ACM Computing Surveys (CSUR)}, 2020.

\bibitem{zhang2017frequency}
C.~Zhang and V.~Prasanna, ``{Frequency Domain Acceleration of Convolutional
  Neural Networks on {CPU-FPGA} Shared Memory System},'' in \emph{Proceedings
  of the 2017 ACM/SIGDA International Symposium on Field-Programmable Gate
  Arrays (FPGA'17)}, 2017, pp. 35--44.

\bibitem{abtahi2018fft}
T.~Abtahi, C.~Shea, A.~Kulkarni, and T.~Mohsenin, ``{Accelerating Convolutional
  Neural Network With {FFT} on Embedded Hardware},'' \emph{IEEE Transactions on
  Very Large Scale Integration (VLSI) Systems}, vol.~26, no.~9, pp. 1737--1749,
  2018.

\bibitem{wang2018c}
S.~Wang, Z.~Li, C.~Ding, B.~Yuan, Q.~Qiu, Y.~Wang, and Y.~Liang, ``{C-LSTM:
  Enabling Efficient LSTM using Structured Compression Techniques on FPGAs},''
  in \emph{Proceedings of the 2018 ACM/SIGDA International Symposium on
  Field-Programmable Gate Arrays (FPGA'18)}, 2018, pp. 11--20.

\bibitem{li2019rnn}
Z.~Li, C.~Ding, S.~Wang, W.~Wen, Y.~Zhuo, C.~Liu, Q.~Qiu, W.~Xu, X.~Lin,
  X.~Qian \emph{et~al.}, ``{E-RNN: Design Optimization for Efficient Recurrent
  Neural Networks in FPGAs},'' in \emph{2019 IEEE International Symposium on
  High Performance Computer Architecture (HPCA)}.\hskip 1em plus 0.5em minus
  0.4em\relax IEEE, 2019, pp. 69--80.

\end{thebibliography}
